\documentclass[a4paper,fleqn,usenatbib]{mnras}
\usepackage{newtxtext,newtxmath}
\usepackage[T1]{fontenc}
\usepackage{ae,aecompl}
\usepackage{longtable}
\usepackage{threeparttable}
\usepackage{natbib}
\usepackage{adjustbox}
\usepackage{graphicx}
\bibpunct{(}{)}{;}{a}{}{,}
\usepackage{booktabs}
\usepackage[T1]{fontenc}
\usepackage{aecompl}
\usepackage{multicol}
\usepackage{multirow}
\usepackage{placeins}
\usepackage{color}
\hypersetup{draft}  

\title[]{Evidence of globular cluster abundance anomalies in the SMC intermediate-age cluster Kron 3}
\author[Salgado et al.]{C. Salgado,$^{1,2}$\thanks{Contact e-mail: \href{mailto:salgado.carolina@gmail.com}{salgado.carolina@gmail.com}}
G. S. Da Costa,$^{1}$
D. Yong,$^{1}$
R. Salinas,$^{3}$
J. E. Norris,$^{1}$
A. D. Mackey,$^{1}$
\newauthor A.F. Marino,$^{4,5}$
A. P. Milone$^{4,6}$ 
\\
$^{1}$Research School of Astronomy and Astrophysics, Australian National University, Canberra, ACT 2611, Australia\\
$^{2}$Departamento de Matem\'atica y F\'isica Aplicadas, Universidad Cat\'olica de la Sant\'isima Concepci\'on, Concepci\'on, Chile\\
$^{3}$Gemini Observatory/NSFs NOIRLab, Casilla 603, La Serena, Chile\\
$^{4}$Istituto Nazionale di Astrofisica - Osservatorio Astronomico di Padova, Vicolo dell'Osservatorio 5, IT-35122, Padua, Italy\\
$^{5}$Istituto Nazionale di Astrofisica - Osservatorio Astrofisico di Arcetri, Largo Enrico Fermi, 5, Firenze, IT-50125\\
$^{6}$Dipartimento di Fisica e Astronomia ``Galileo Galilei'', Universit\'a di Padova, Vicolo dell'Osservatorio 3, I-35122, Padua, Italy\\
}

\date{Accepted XXX. Received YYY; in original form ZZZ}

\pubyear{2021}

\begin{document}
\label{firstpage}
\pagerange{\pageref{firstpage}--\pageref{lastpage}}
\maketitle

\begin{abstract}
Using spectra obtained with the VLT/FORS2 and Gemini-S/GMOS-S instruments, we have investigated carbon, nitrogen and sodium abundances in a sample of red giant members of the Small Magellanic Cloud star cluster Kron~3.  The metallicity and luminosity of the cluster are comparable to those of Galactic globular clusters but it is notably younger (age $\approx$ 6.5 Gyr). We have measured the strengths of the CN and CH molecular bands, finding a bimodal CN band-strength distribution and a CH/CN anti-correlation.  Application of spectrum synthesis techniques reveals that the difference in the mean [N/Fe] and [C/Fe] values  for the CN-strong and CN-weak stars are $\Delta<$[N/Fe]$>$ = 0.63 $\pm$ 0.16 dex and \mbox{$\Delta<$[C/Fe]$>$ = --0.01 $\pm$ 0.07 dex} after applying corrections for evolutionary mixing. We have also measured sodium abundances from the Na~D lines finding an observed range in [Na/Fe] of $\sim$0.6 dex that correlates positively with the [N/Fe] values and a $\Delta<$[Na/Fe]$>$ = 0.12 $\pm$ 0.12 dex.  While the statistical significance of the sodium abundance difference is not high, the observed correlation between the Na and N abundances supports its existence.  The outcome represents the first star-by-star demonstration of correlated abundance variations involving sodium in an intermediate-age star cluster.  The results add to existing photometric and spectroscopic indications of the presence of multiple populations in intermediate-age clusters with masses in excess of $\sim 10^5$ M$_{\odot}$.  It confirms that the mechanism(s) responsible for the multiple populations in ancient globular clusters cannot solely be an early cosmological effect applying only in old clusters.

\end{abstract}


\section{Introduction}
\label{s_intro3}

The traditional picture of globular clusters as simple stellar populations has suffered radical changes over the past few decades. Photometric and spectroscopic  studies have demonstrated that  globular clusters host multiple populations (MPs) of stars. These MPs are characterised by  star-to-star abundance variations in the light elements (C, N, O, Na, Mg and Al) and in helium,  although otherwise the stars are generally homogeneous in composition as regards the heavier elements such as Ca, Fe and the $s$-process elements.  Star-to-star variations in these heavier elements are found, however, 
in a number of predominantly luminous globular clusters (see, for example, \citet{2016IAUS..317..110D, marino2021spectroscopy}, and the references therein).

In Galactic globular clusters (GGCs) the fraction of stars showing the abundance anomalies, i.e., stars depleted in carbon, oxygen and magnesium, and enhanced in nitrogen, sodium and aluminium (and also helium), ranges from $\sim$35\% to 90\% with the larger fractions occurring in more massive clusters \citep{ milone2017hubble}.  The cluster stars  that do not show the anomalies have abundance ratios comparable to field halo stars of the same metallicity \citep[e.g.,][]{OsbornW1971,cottrell1981, norris1981abundance,kraft1994abundance,gratton2001and,gratton2004abundance,salaris2006primordial, carretta2005abundances, carretta2009anticorrelation, renzini2015hubble, carretta2016spectroscopic, bastian2018multiple, gratton2019globular}.  The origin of the abundance variation likely lies with H-burning at sufficiently high temperatures. To alter the abundance of light elements  a hot H-burning
environment is required to allow the simultaneous action of p-capture reactions in the CNO, NeNa and MgAl chains \citep{denisenkov1989,langer1993, prantzos2007light}. The high temperatures  required for this process are $\geq {20 \times 10}^6$~K and $\geq {70 \times 10}^6$~K for the NeNa and MgAl cycles, respectively. However, these temperatures cannot be reached in the cores of current GGC stars, which necessarily implies a multigeneration scenario in the cluster formation process. Therefore, the stellar population of GGCs can be thought of as a mix of  first  generation stars (primordial) and second generation stars (polluted). We note that we use of terms ``first'' and ``second'' generation as population labels; its use does not necessarily imply a requirement for a time interval between the formation of the populations.  For example, scenarios exist that suggest all the stars formed at the same time with the chemical anomalies arising from accretion of polluted material \citep[e.g.,][]{bastian2013early, gieles2018concurrent}. Several candidates have been proposed as the origin of the pollution: intermediate-mass asymptotic giant branch (AGB) stars~\citep{cottrell1981,dantona1983,ventura2001} and/or super-AGB stars \citep{pumo2008super,ventura2011,d2016single}, supermassive stars \citep{denissenkov2014supermassive,denissenkov2015primordial, gieles2018concurrent}  and fast rotating massive stars (FRMS) \citep{norris2014, maeder2006, decressin2007fast, decressin2009cno}. See also \citet{gratton2012multiple}, \citet{renzini2015hubble}, \citet{bastian2018multiple} and \citet{gratton2019globular} for critical reviews of the multiple-population formation scenarios.

An important point to note is that variations of abundances have been  consistently found among GGCs, but this phenomenon is rarely seen in the halo field population \citep{martell2011building, schiavon2017chemical}. As for globular clusters distinct from those of the Milky Way, \citet{larsen2014nitrogen}, through measurements of nitrogen abundances in red giants based on HST multi-band photometry, has shown that four metal poor clusters in the Fornax dSph exhibit the chemical anomalies. In addition, \citet{letarte2006vlt} has found indications for the presence of the anomalies from spectroscopy of nine stars in three globular clusters associated with  Fornax dSph galaxy. Old and metal-poor star clusters in the Large Magellanic Cloud (LMC), like the GGCs, also possess Na-O and Mg-Al anti-correlations \citep{mucciarelli2012}. Further, several studies \citep[e.g.,][]{carretta2010properties, milone2014helium, salinas2015no, lagioia2019role, milone2020multiple} have suggested that cluster mass plays an important role in generating the MPs. However, studies such as \citet{mucciarelli2008chemical}, \citet{milone2020multiple} and \citet{ martocchia2021nitrogen} have shown that the abundance anomalies are not obviously present in LMC and Small Magellanic Cloud (SMC) star clusters with ages less than $\sim$2 Gyr, despite these clusters having present-day masses comparable to the present-day masses of GGCs that possess abundance anomalies.  While the best comparison would be with initial masses, these are uncertain especially for the GGCs given the potential variety of dynamical evolution effects from interactions with the Milky Way potential.  Nevertheless, the results may indicate an age dependence for the occurrence of multiple populations \citep[e.g.,][]{martocchia2017age}.

Since the LMC does not contain any massive clusters older than $\sim$3 Gyr and younger than the $\sim$13 Gyr age of the GGCs, the luminous star clusters of the SMC become the obvious targets to investigate this question.  The clusters cover a substantial age range and their properties, such as present-day masses, length scales and metallicities are very similar to those of the GGCs except for the younger ages.  In particular, photometric and spectroscopic studies of luminous SMC star clusters with ages exceeding 2 Gyr generally show evidence for the presence of MPs.  

NGC~121 is the only `classical' globular cluster that belongs to SMC with an age of $\sim$11 Gyr \citep{glatt2008age}. Using HST data, \citet{niederhofer2016search2} found that NGC~121 has a split red giant branch, analogous to that seen for many GGCs. However, the fraction of enriched stars (N rich and C poor) is $\sim$32 percent, which is substantially less than the typical value for Milky Way GCs of comparable present-day mass. These results are consistent  with  \citet{dalessandro2016multiple}, that characterized the populations of this cluster with a combination of optical and near-UV HST photometry and  ESO-VLT/ FLAMES  high-resolution spectroscopy.  They reported the detection of  multiple populations in NGC 121 stating, in agreement with \citet{niederhofer2016search2}, who the cluster is  dominated by first generation stars (more than 65\%). On the other hand, based on a chromosome-map analysis  \citet{milone2020multiple} estimate that the fraction of first generation stars in NGC~121 is 52 $\pm$ 3 percent, a value that remains somewhat higher than the first generation fractions for Milky Way GCs with similar present-day masses.

Nine star clusters in the Magellanic Clouds with ages between $\sim$1.5 and $\sim$11 Gyr  and masses similar to that of GGCs were investigated in \citet{martocchia2017age}. The photometric analysis confirmed the presence of MPs in all the clusters studied older than 6 Gyr. Further, the study revealed photometric evidence for MPs on the red giant branch of NGC~1978, an LMC cluster of age $\sim$2.3 Gyr  \citep{martocchia2018search}.  This result has been confirmed  by an analysis of the strengths of CN- and CH-bands in the spectra of 24 member stars \citep{martocchia2021nitrogen}; see also the recent results of \citet{Li2021msstars}. 
\citet{martocchia2021nitrogen} has also found a significant intrinsic spread in CN in 21 members of NGC 1651 (age $\sim$2 Gyr), a signal of the presence of MPs. These findings reveal NGC 1651 as apparently the youngest cluster to host abundance anomalies. On the contrary, no detection of MPs in clusters younger than \mbox{$\sim$1.7 Gyr.} were reported \citet{martocchia2021nitrogen}.
Further, \citet{milone2020multiple} (see also \citet{lagioia2019helium}) have constructed chromosome-map diagrams for  11 GCs of Large and Small Magellanic Clouds with different ages, detecting MPs in five clusters. The fraction of first population stars found ranges between 50\% and 80\%, which is, in at least some cases, significantly higher than the fraction observed in GGCs with similar present-day masses (for more details also see \citet{dondoglio2021multiple}).

The strengths of the CN and CH molecular bands in optical spectra are excellent indicators of abundance variations. In GGCs, CN- and CH- band strengths generally show an anti-correlation coupled with a bimodal or multi-modal distribution of CN-band strength \citep[e.g.,][]{norris1981abundance, norris1984anticorrelation, pancino2010}. The variation of CN- and CH-band strengths are a consequence of differences in carbon and nitrogen abundances. In addition, the anti-correlation between sodium and oxygen abundances seen in stars at all evolutionary phases in GGCs \citep[e.g.,][]{ gratton2001and, carretta2009anticorrelation, carretta2016spectroscopic} emphasises the complexity of the GGCs populations. The phenomenon is also manifest in GGCs through the correlation between CN-band strengths and sodium, where CN-strong stars (enhanced in nitrogen and depleted in carbon) are also enhanced in sodium \citep{cottrell1981,  norris1985sodium, salgado2019investigation}.

Star-to-star N/C variations in cluster red giant stars are likely the result of MPs, but their existence is not sufficient on their own to assert that their origin is necessarily the same as for those seen in GGCs.  This is because evolutionary mixing can occur on the red giant branch (RGB) reducing the carbon abundance and raising the nitrogen abundance through CN-cycle processing \citep{salaris2015post, salaris2020photometric, cadelano2021expanding}. Sodium, however, is unaffected by such mixing and thus Na abundance variations are an unambiguous indicator of the presence of MPs.  Sodium abundance variations are, however, not easily established.  For example, \citet{saracino2020leveraging} and \citet{martocchia2020leveraging} have used a combination of Hubble Space Telescope (HST) and ESO VLT MUSE observations to reveal the presence of Na abundance variation in intermediate-age massive star clusters. \citet{saracino2020leveraging} found a Na difference 
of $\Delta<$[Na/Fe]$>$ = 0.07 $\pm$ 0.01 between the N-rich and N-poor stars of NGC 1978 (age $\sim$2 Gyr), while \citet{martocchia2020leveraging} found a  mean abundance variation of $\Delta <$[Na/Fe]$>$ = 0.18 $\pm$ 0.04 dex for NGC 416 (age $\sim$6.5 Gyr) and $\Delta <$[Na/Fe]$>$ = 0.24 $\pm$ 0.05 dex for Lindsay 1  (age $\sim$7.5 Gyr): the population that is enhanced in N also enhanced in Na.

The aim of this work is to study abundance variations in the light elements, like those seen in GGCs, in a massive star cluster of SMC. We study  Kron~3, a cluster with an age of $\sim$6.5 Gyr and a metallicity and luminosity comparable to GGCs \citep{glatt2008age}. Specifically, in this paper we investigate the distribution of CN- and 
CH-band strengths in red giant branch stars looking for the anti-correlation and bimodality seen in GGCs. However, because of possible evolutionary mixing on the RGB, a CN -- CH anti-correlation is not sufficient in itself to indicate the presence of anomalies. Consequently, we also explore the strengths of the NaD lines, as an additional constraint, looking for a Na/CN correlation similar to the correlation seen in GGCs. C, N and Na abundances are estimated through the application of synthetic spectra calculations.

The paper is arranged as follows. The observations, the data reduction, and the definition of indices and their measurement are presented in Section \ref{s_obs3}.  Section \ref{membership_3} discusses the criteria used for cluster membership determination and the results are presented and discussed in Section \ref{results cnchna_smc} and Section \ref{disc_3}.  The outcomes  are summarized in Section \ref{concl_3}.

\section{Observations and data reduction}
\label{s_obs3}

Our data set consists of spectra of red giant stars from the SMC globular cluster Kron~3. Table \ref{cluster_parameter} lists properties of the cluster. The 31 stars observed were chosen to lie on the cluster RGB between the tip and approximately 0.5 magnitude above the level of the cluster red horizontal branch. The faintest star in the observed sample has \mbox{V $\simeq$ 19}. 

\begin{figure*}
  \centering
  \includegraphics[width=13cm]{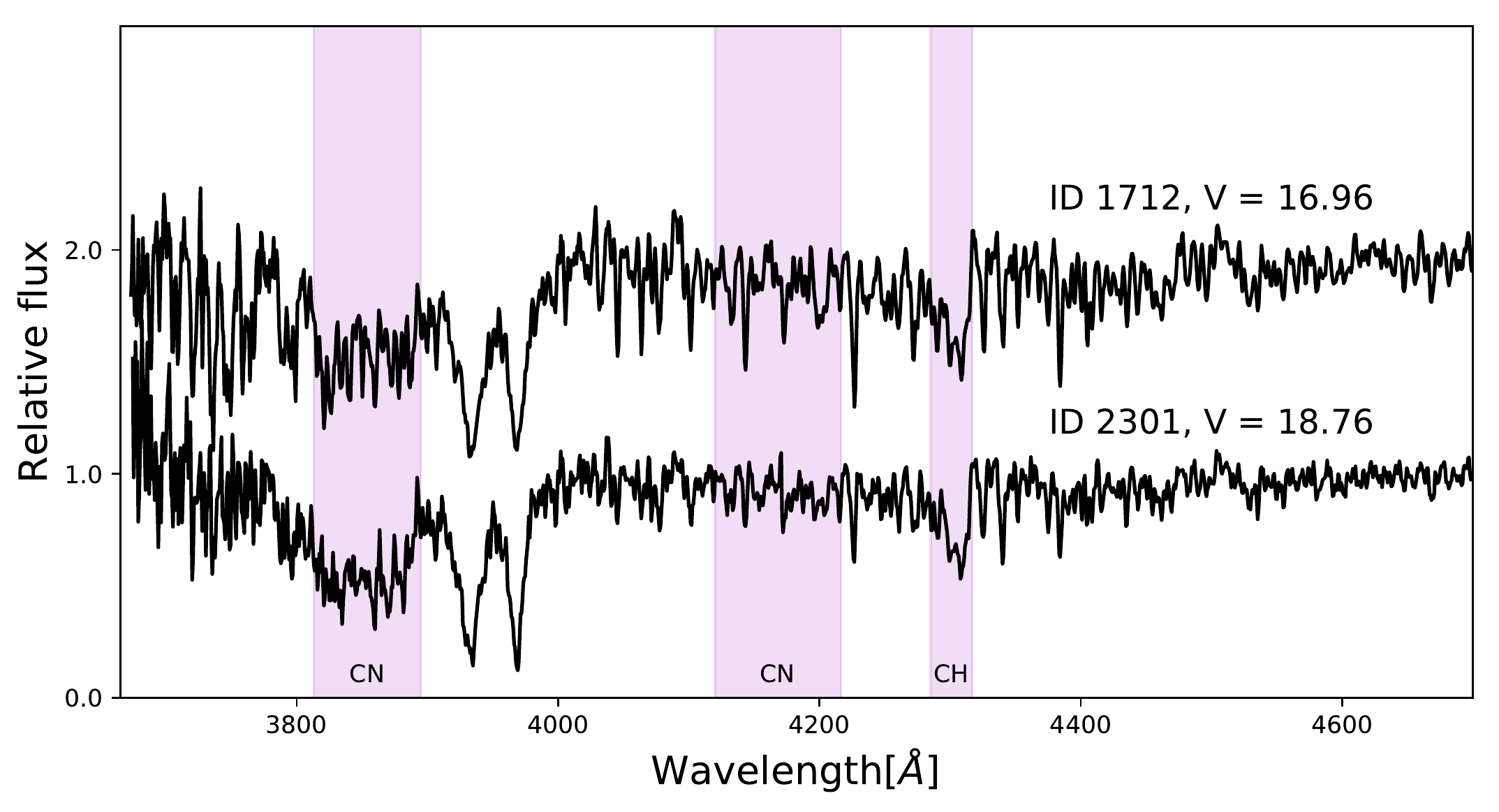}
  \caption{Examples of a subsection of the full continuum-normalized ESO/FORS2 spectra of Kron~3 target stars with different magnitudes. The upper spectrum was vertically shifted by 1.0 to avoid the overlap. IDs and magnitudes are indicated. The location of the CH-, at $\lambda \sim$ 4300 \AA \ and the CN-, at $\lambda \sim$ 3883  \AA \  and $\lambda \sim$ 4215 \AA \, bands  are shown by the shaded regions.} 
\label{cnch_spectra_SMC}
\end{figure*}

\begin{figure*}
  \centering
  \includegraphics[width=13cm]{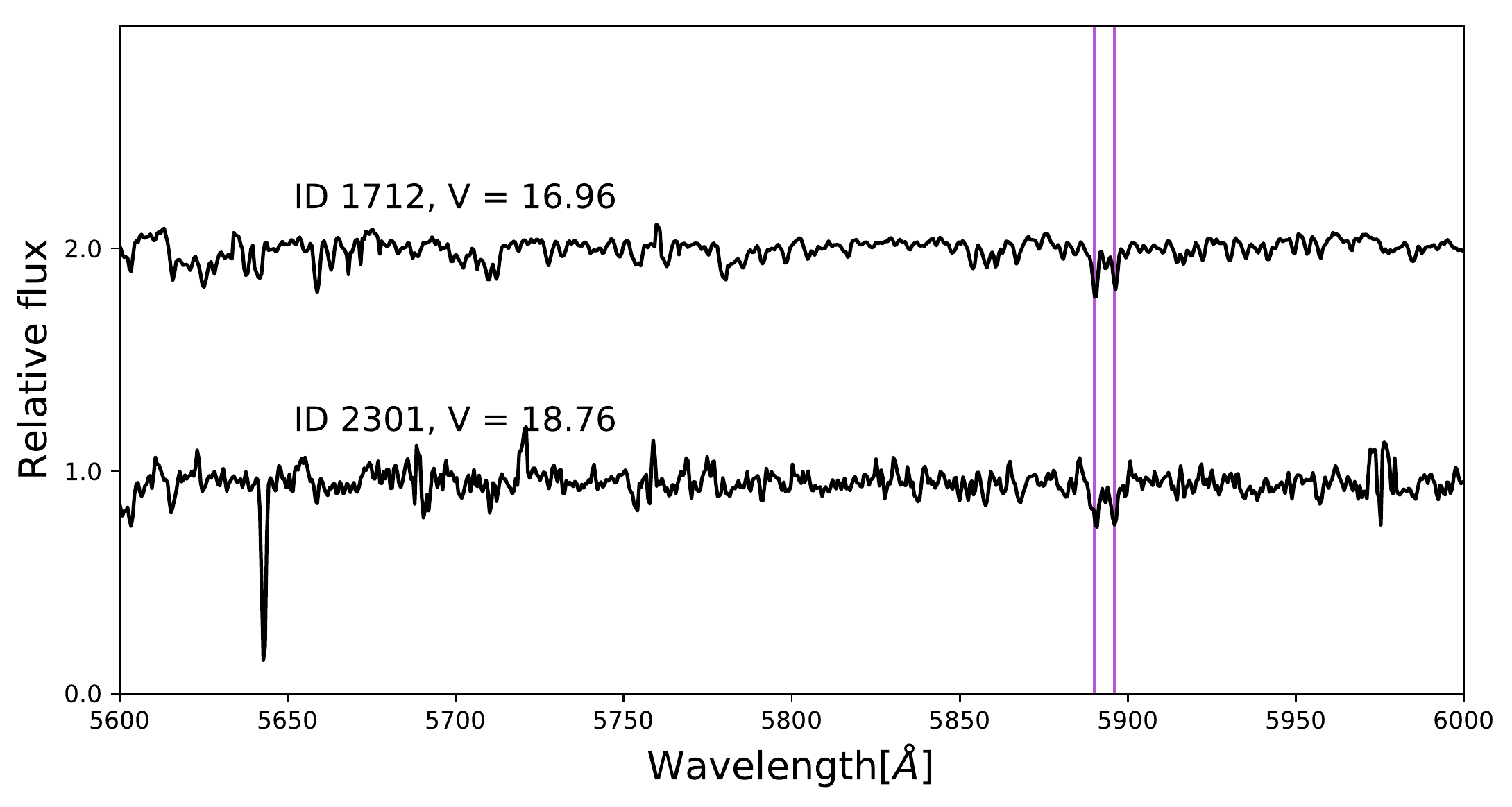}
  \caption{Examples of a subsection of the full continuum-normalized Gemini/GMOS-S spectra of Kron~3 target stars with different magnitudes. The upper spectrum was vertically shifted by 1.0 to avoid the overlap. IDs and magnitudes are indicated. The vertical magenta lines indicate the position of the NaD lines.} 
\label{na_spectra_SMC}
\end{figure*}

We have observations that include the CN- and CH-bands at UV and blue wavelengths, as well as spectra that cover the D-lines of sodium at red wavelengths. For the blue sample, the cluster was first pre-imaged with FORS2 in imaging mode in the B and V filters: 2 pairs of short 
exposure B and V images were obtained with a small spatial offset between the first and the second B, V pair.   The raw data frames were processed through the standard FORS pipeline and aperture 
photometry performed on each pair and the values averaged to determine a colour-magnitude diagram (CMD). The instrumental FORS2 magnitudes and colours were then converted to the standard B, V system using zero points, transformation equations and extinction coefficients
available from the ESO archive.  The CMD was then used to select stars that lay on the RGB; those stars were then input to the MOS slit allocation process and the best configurations of MOS-slits and candidates generated.
The blue data (obtained under ESO Program 095.D-0496) were collected between June and September 2015 at the VLT, Cerro Paranal, Chile, using FORS2 in MOS mode with the 1200~B + 97 grism. The seeing, sky transparency and airmass met the requested conditions (1.0'', clear, $\leq$ 1.7 respectively). The instrument setup covers wavelengths from 3700 to 5150 \AA \ at a scale of 0.72 \AA /pix and with a resolution of 3 \AA \ for 1'' slits. With this configuration, the observed spectra cover the G-band (CH $\sim$4300 \AA) and the violet ($\sim$3880 \AA)\ and blue ($\sim$4215 \AA) CN-bands. Two different MOS configurations were observed with integrations of 2$\times$1300 sec; the two exposures  allow cosmic-ray rejection. The VLT/FORS2 spectra were reduced, extracted and wavelength-calibrated with the ESO Recipe Flexible Execution Workbench (Reflex)\footnote{\small{https://www.eso.org/sci/software/esoreflex/}}. 
The signal-to-noise of the spectra was estimated at $\lambda \sim$ 4200 \AA\,  but due to the difficulty in setting the continuum, especially in the cooler stars, we used the FORS exposure time calculator to confirm the results. The S/N at $\lambda$4200 \AA\, for a star with magnitude V = 19 is $\sim$13, while the S/N is higher for the brighter stars ($\sim$60 for a star with magnitude \mbox{V = 16.8}).  These values match satisfactorily the S/N seen in the spectra. Examples of the blue (and red, see below) spectra for stars with different magnitudes are shown in Fig.\ \ref{cnch_spectra_SMC} and Fig.\ \ref{na_spectra_SMC}.

A subset of the stars observed with FORS2 were also observed with the Gemini multi-object spectrograph (GMOS-S;  \citet{hook2004gemini}) under the Gemini program GS-2017B-Q-57. Because of the smaller field-of-view of GMOS-S compared to FORS2, it was not possible to observe all stars from the blue sample. The observations employed the B~1200 grating centered at 
$\lambda$5350 \AA \ giving a wavelength coverage from $\sim$4600 to $\sim$6100 \AA \ for a centrally located star at a spectral resolution of 3.1 \AA.  The  wavelength coverage was chosen to include the NaD lines at $\sim$5900 \AA.  One  mask was observed. In order to compensate for the inter-chip gaps, three integrations were made at slightly different central wavelengths (5250, 5350 and 5450 \AA), which were then combined into a single spectrum.  Masks were designed using the Gemini MOS mask preparation software (GMMPS, v 1.4.5)\footnote{\small{https://gmmps-documentation.readthedocs.io}}. GMMPS uses input catalogues with target positions and priorities, and optimizes the number of targets per mask, by applying vertical offsets along the slits. The reduction of the red spectra was made using standard IRAF/Gemini software. The signal-to-noise in the GMOS spectra was estimated at \mbox{$\lambda$6000 \AA\,} using the Gemini GMOS-S integration time calculator. Three 1350 sec exposures generate S/N \mbox{$\sim$45} per spectral pixel for a  star of magnitude V = 19 at \mbox{$\lambda$ 6000 \AA\,}, which again matches reasonably with the S/N seen in the observed spectra of stars of similar magnitude, which are the faintest in our sample.
Details of the observations are shown in Table \ref{Tabla_obs3}.

%
%
%
%

\begin{table}
\caption{Cluster parameters.}
\centering
\renewcommand{\TPTminimum}{\linewidth}
\begin{adjustbox}{max width=8.5cm}
\setlength{\tabcolsep}{12pt} 
\renewcommand{\arraystretch}{1.3} 
\begin{tabular}{l c c c } 
\hline 
\hline
\normalsize{\textbf{Cluster name}} & \normalsize{\textbf{Mass $^{1}$} } & \normalsize{\textbf{Age $^{2}$ }} & \normalsize{\textbf{[Fe/H] $^{3}$} }  \\ [0.5ex] 
\small{} & \small{\textbf{10$^{5}$~M$_{\odot}$} } & \small{\textbf{Gyr}} & \small{}  \\ [0.5ex] 
\hline 

Kron~3		&	5.8	& 6.5 $\pm$ 0.5	& --1.08 $\pm$ 0.12	\\

\hline 
\end{tabular}
\end{adjustbox}
\begin{tablenotes}
\item [1] (1) \citet{glatt2011present} (Mass determined using a Kroupa-like IMF). \citet{song2021dynamical} found a mass of 0.77$\times$10$^{5}$~M$_{\odot}$ from a dynamical study.
\item [2] (2) \citet{glatt2008age}
\item [3] (3) \citet{da1998ii}
\end{tablenotes}
\label{cluster_parameter}
\end{table}

\begin{table}
\caption{Details of the observations.}
\renewcommand{\TPTminimum}{\linewidth}
\centering 
\begin{adjustbox}{max width=8.5cm}
\setlength{\tabcolsep}{15pt} 
\renewcommand{\arraystretch}{1.3} 
\begin{tabular}{l c c } 
\hline 
\hline
\normalsize{} & \normalsize{\textbf{VLT/FORS2} }   & \normalsize{\textbf{GMOS-S}}    \\ [0.5ex] 
\small{} & \small{\textbf{Blue spectra} } & \small{\textbf{Red spectra}}     \\ [0.5ex] 
\hline 

Dispersion	element  		&	1200 B + 97								& B 1200									\\
Resolution (\AA)			&	3										& 3.1										\\
Observed stars				&	33										& 18	\\
Wavelength coverage (\AA)	&	3700 -- 5150								& 4600 -- 6100 								\\

\hline 

\end{tabular}

\label{Tabla_obs3} 
\end{adjustbox}
\end{table}

\subsection{Feature strengths}
\label{strengths3}

Once the blue and red spectra were reduced, extracted, and wavelength-calibrated, they were continuum-normalised by fitting a low-order polynomial function to the stellar continuum.  Velocity corrections, using the observed velocity derived from each spectrum, were then applied to shift the spectra to rest wavelength. The procedure used for the determination of the observed velocities is explained in detail in Section \ref{membership_3}.
The  measurement of CH- and CN- band strengths was then made via numerical integration to calculate the indices S(3839), S(4142) and W(G) \citep{norris1981abundance, norris1979cyanogen, norris1984anticorrelation}, using the identical procedures to those employed in \citet{salgado2019investigation}. In addition, the equivalent widths (EW) of the NaD sodium absorption lines at $\sim$5889 \AA \ and $\sim$5895 \AA \ were determined via gaussian fitting using standard routines in IRAF; this is also the same process as used in \citet{salgado2019investigation}.

\section{Membership determination}
\label{membership_3}

An accurate cluster membership determination is crucial to interpret correctly the strengths of the CH- and CN-bands, NaD absorption line strengths, and the corresponding C, N and Na abundances. Consequently,  three criteria were utilised to determine the cluster membership: (1) radial velocity (RV) measurements and distance from the cluster centre; (2) measurements of equivalent widths of Ca{\sc ii}~K lines; and (3) location on the RGB in the CMD. These criteria do not act independently, but rather go together as explained below.

We consider as a starting point the distance of the targets from the cluster centre. This criterion used as an initial guide the limiting radius of the cluster defined by a fit of a King surface-desnity profile \citep{king1962structure}, which is given in \citet{glatt2009structural} as 180 $\pm$ 38 arcsec for Kron~3.  However, we decided not to be so strict initially, and adopted a limiting radius for possible candidate members as 250 arcsec (i.e., approximately +2$\sigma$ from the \citet{glatt2009structural} limiting radius).

We then considered the radial velocities (RVs) of the target stars.  The RVs  were measured using the IRAF task \textit{fxcor}, which is based on the cross-correlation method presented in \citet{tonry1979survey}. The radial velocities used 
were measured only from FORS2 spectra, and the correlation was computed over the 4000 to 4800 \AA \ range. We did not use the radial velocities obtained from the GMOS-S spectra because the wavelength calibration is uncertain, given that the arc lamp exposures were taken on the day following the observations. We adopted as a template a synthetic spectrum of a typical red giant with stellar parameters comparable to those of the program stars. According to \citet{morse1991measurement} and \citet{moni2011lack}, in terms of RVs the correct choice of template is not crucial because the discrepancy between the template and the object spectra only increases the uncertainties in the measurements, but does not introduce systematic errors. In order to check the consistency of the RVs they were measured separately from the two sets of observations for each FORS2 MOS setup, and then the velocity for each star calculated as the weighted arithmetic mean of the 2 measures using the \textit{fxcor} errors as weights. Heliocentric corrections then were applied, and the mean and standard deviation ($\sigma$) of the set of velocities calculated.  Stars that lay more than $\pm2.5\sigma$ from the mean were flagged as candidate non-members.

\begin{figure}
  \centering
  \includegraphics[width=8.5cm]{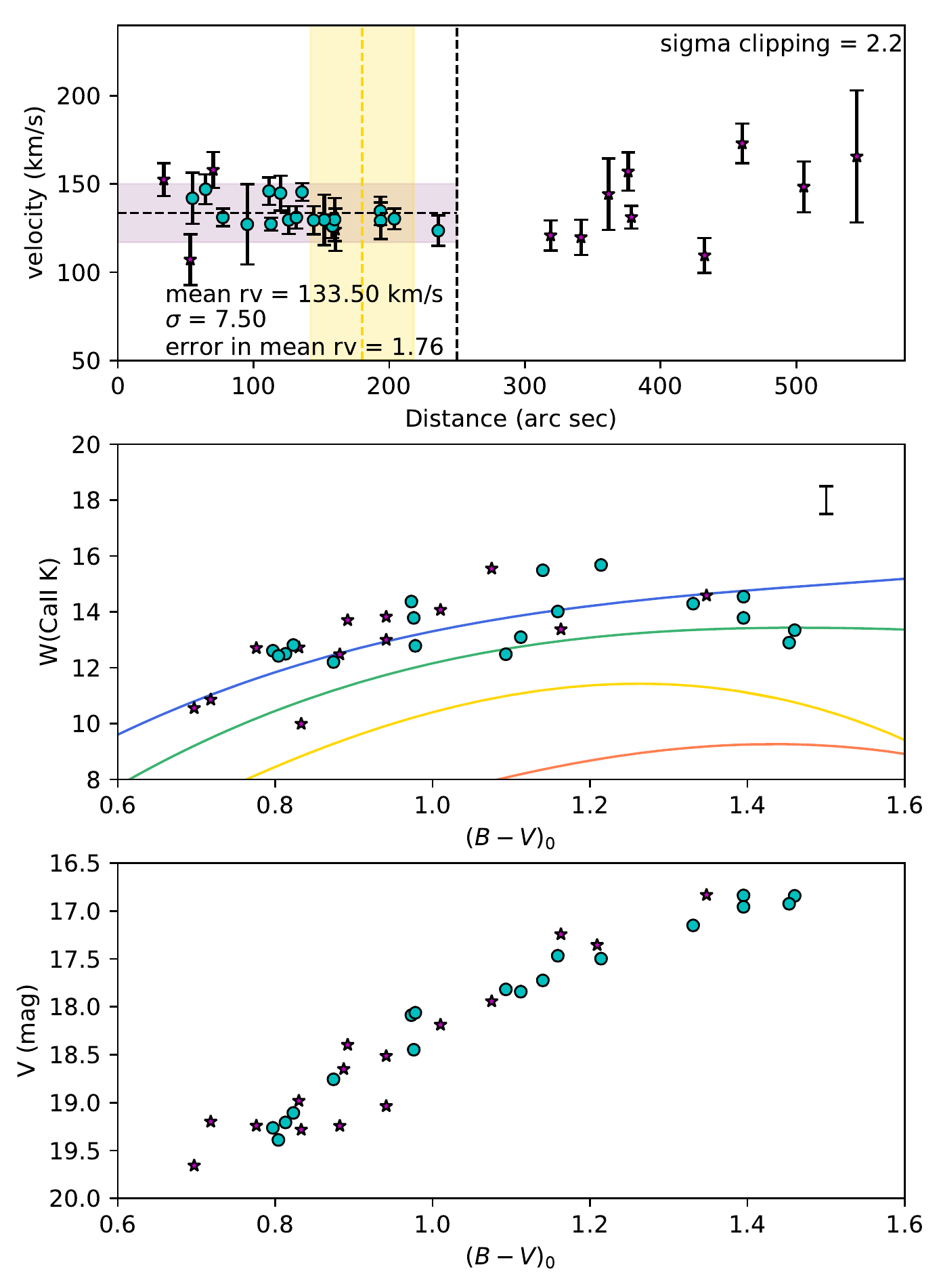}
  \caption[The three criteria adopted to define the membership in Kron~3.]{The three criteria adopted to define the membership in Kron~3. In all panels filled aqua circles represent the adopted members and the magenta star symbols are the non-members. The upper panel represents radial velocities against distance from the centre of the cluster. Error bars from the cross-correlation procedure are shown.  The horizontal dashed line represents the mean radial velocity obtained from the FORS2 spectra. The magenta shaded region represents an area centred on the mean velocity $\pm$ 2.2$\sigma$. The yellow dashed line and shaded yellow region represent the estimated cluster tidal radius and its uncertainty from \citet{glatt2009structural}, respectively. The vertical dashed line shows the adopted cutoff radius for cluster members. The central panel shows the equivalent width of the Ca{\sc ii}~K line for the observed stars against colour (B-V)$_{0}$. A typical $\pm$1$\sigma$ index error value is shown in the top right corner. The blue, green, yellow and orange solid lines represent the calibrated models of expected varition in the EW of Ca{\sc ii}~K line with $(B-V)_{0}$ colour for [Fe/H] = --1.0, --1.5, --2.0 and --2.5, respectively \citep{beers1990estimation}.   The lower panel shows the position of the  observed stars in the CMD.}   
\label{vel_smc_all}
\end{figure}

The next step involved metallicity determinations, as it is reasonable to assume that the Kron 3 members will all have the same metallicity (see Table \ref{cluster_parameter}). Specifically, we have used Ca{\sc ii}~K line strengths  to investigate the metallicities of the observed stars and thus their cluster membership status. The EW of Ca{\sc ii}~K line was measured on each spectrum and we used the calibration of Ca{\sc ii}~K line line strength with metallicity of \citet{beers1990estimation} to determine metallicity estimates.  In particular, \citet{beers1990estimation} describes a method for estimating metallicity ([Fe/H]) from Ca{\sc ii}~K line strengths in spectra with 1 -- 2 \AA \ resolution covering the wavelength range 3700 -- 4500 \AA \  via a model calibration of the expected variation of the Ca{\sc ii}~K line equivalent width as a function of (B-V)$_0$ colour and metallicity. The appropriate \citet{beers1990estimation} calibration to be employed depends on the strength of the Ca{\sc ii}~K line, which is measured in different wavelength bands depending on its strength. For our stars the line strengths suggest measurement in an 18 \AA \ band, with the strength denoted by K18. The polynomial coefficients from \citet{beers1990estimation} for K18 line strengths were then used, with the (B-V)$_0$ values, to generate a metallicity estimate for each star. The mean and rms of the deviations in W(Ca{\sc ii}~K) from the [Fe/H] = --1.0 calibration line at the (B-V)$_0$ of each likely member star was then calculated. The rms is $\sim$1 \AA, which, under the assumption of no intrinsic variation in the cluster star [Fe/H] values, provides an estimate of the uncertainty in the W(Ca{\sc ii}~K) values.  We can convert this error in W(Ca{\sc ii}~K) into a error in [Fe/H] using the separation between the lines for [Fe/H] = --1.0 and [Fe/H] = --1.5 at a colour of (B-V)$_0$ = 1.1, resulting in an estimated $\sigma$([Fe/H]) of $\sim$0.45 dex for each individual star. This value indicates that 
the  W(Ca{\sc ii}~K) diagram does not provide strong discrimination between members and non-members, though it is evident from Figure \ref{vel_smc_all} that some stars can be excluded from cluster membership by this criterion.  We note also that the mean deviation of the likely cluster members from the [Fe/H] = --1.0 line is only --0.02 \AA \, so that the estimated abundance for Kron~3 is [Fe/H] $\approx$ --1.0, entirely consistent with other abundance estimates for this cluster.

We also used the location in the CMD of the observed stars as a further membership criterion: stars with larger deviations from the mean RGB locus were flagged as possible non-members.

In practice all of these criteria were used iteratively to arrive at a final list of 18 probable cluster members.  For example, the sigma-clipping of the radial velocities was ceased at $\sigma$ = 2.2 after several iterations, as at that value the remaining stars met all the criteria for cluster membership.  The outcome is shown the panels of Figure \ref{vel_smc_all} where it is evident that radial velocity and distance from the cluster centre are the strongest membership discriminants, however, metallicity and CMD location provide additional information in each iteration.
For the 18 stars considered as probable Kron~3 members, the mean radial velocity (RV)  is 133.5 km s$^{-1}$ and the standard error of mean $\varepsilon$ is 1.7 km s$^{-1}$.  Our value is compatible with those found by \citet{parisi2015ii} (RV= 135.1 km s$^{-1}$),   \citet{hollyhead2018kron} (RV= 135.9 km s$^{-1}$), and with the precise determination of \citet{song2021dynamical} (RV= 132.7 $\pm$ 0.4 km s$^{-1}$).

\begin{figure}
\centering
\includegraphics[width=7.5cm]{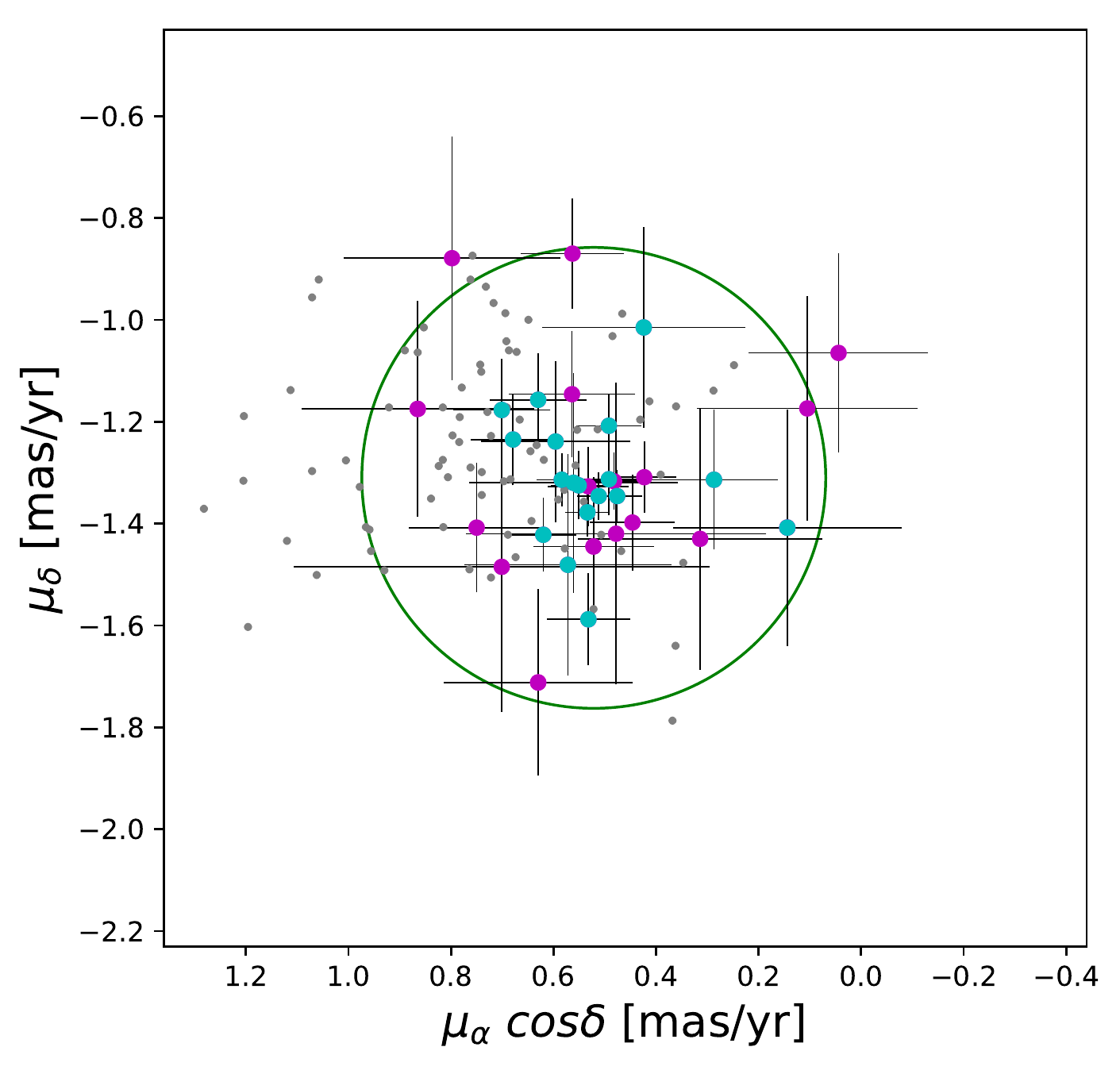}

\caption[Gaia EDR3 proper motions for the stars observed in Kron~3]{Gaia EDR3 proper motions for the stars observed in Kron~3 are shown. Cyan dots represent the adopted cluster members, while magenta dots represent the non-member stars. The green circle has a centre determined as the mean pm in RA  and the mean pm in  DEC of the candidate members with a  radius of 2.5$\times \sigma$,  where $\sigma$ is defined as  
$\sqrt{\sigma_{pm\/RA}^2 + \sigma_{pm\/DEC}^2}$, where $\sigma_{pm\/RA}$ and $\sigma_{pm\/DEC}$ are the mean standard deviations of the proper motions in each coordinate for the candidate member stars. Grey dots represent the proper motions of all bright (17.5$<$G$<$18.5) Gaia EDR3 stars in a radius of 3 arcmin from the Kron~3 centre. }

\label{proper_motion_SMC}
\end{figure}

For completeness we show in Figure \ref{proper_motion_SMC} the proper motions from Gaia EDR3 \citep{2020yCat.1350....0G} for the stars observed. The centre of the green circle is the mean proper motion in RA and the mean proper motion in DEC for the probable cluster members while the radius is 2.5 $\times$ $\sqrt{\sigma_{pm\/RA}^2 + \sigma_{pm\/DEC}^2}$, where $\sigma_{pm}$ corresponds to the mean standard deviation in each coordinate of the proper motions of the members.  All probable cluster members have consistent proper motions.

Table \ref{tabla_total_index_SMC} and \ref{tabla_total_index_SMC_nonmember} list the IDs, positions, photometry, radial velocities and distances from the cluster centre for both the adopted  members and non-members respectively.

\section{Results} 

\label{results cnchna_smc}

	\subsection{CH and CN}
	\label{results_CH_CN}

In \citet{salgado2019investigation} we explained how the abundance anomalies are  characterized through the analysis of red giant spectra in the vicinity of the CN- and CH-bands. Previous work has shown that stars belonging to Galactic globular clusters appear as two well separated groups of CN-weak and CN-strong stars, and that CN-strong stars are CH-weak and vice-versa \citep[e.g.,][]{pancino2010,kraft1994abundance}. 
In this work we investigate if these GGCs anomalies are also seen in the intermediate-age Small Magellanic Cloud cluster Kron~3. For this purpose we have measured the indices W(G), S(3839) and S(4142), as explained in section \ref{strengths3}, using the FORS2 spectra for the Kron~3 members.

Table \ref{tabla_total_index_SMC}  gives the values of the indices as well as the sodium line strengths from the GMOS-S spectra (see $\S$\ref{sodium_smc} below).  
Figures \ref{fig_panel_SMC_3839} and \ref{fig_panel_SMC_4142_wg} show the measured band strengths against $V-V_{HB}$ for the Kron~3 members. The V$_{HB}$ value used was 19.5 and it was determined from the cluster CMD generated with the FORS2 pre-imaging photometry.

Figure \ref{fig_panel_SMC_3839}  shows the index S(3839), which was used to classify, via visual inspection,  the cluster stars as either  CN-weak or CN-strong.  The plot reveals that CN-strong stars are well separated from CN-weak population of the cluster, with an index difference of about  $\sim$0.3 mag.
The dashed line, which is a least-squares fit to the data for the CN-weak stars, has a slope of --0.12 with a rms about the fit of 0.05 mag. Under the assumption that at fixed luminosity there is no intrinsic scatter in the CN-weak band strengths, the rms can be used as an error estimate for the S(3839) values.  The dotted lines in the Figure show $\pm$ 2$\times$rms above and below the mean line; stars that lie above the upper dotted line are the CN-strong objects.  There are 6 such stars in our Kron~3 sample and they are plotted as filled symbols in this and subsequent plots.

Figure \ref{fig_panel_SMC_4142_wg} follows the same methodology as for Figure \ref{fig_panel_SMC_3839} for the indices S(4142) and W(G).  In particular, the upper panel of Figure \ref{fig_panel_SMC_4142_wg} confirms that the stars classified as CN-strong from S(3839) also generally have larger values of S(4142) compared to the CN-weak stars.  The slope of the dashed line fitted to the CN-weak stars is --0.07,  the rms about the fit is 0.03 mag,  and the dotted lines are again $\pm$2 $\times$ rms.  We used the rms value as the error for the S(4142) values.  The lower panel of Figure \ref{fig_panel_SMC_4142_wg} reveals that the
CN-strong and CN-weak stars are not as clearly separated in this index as they are in the S(3839) and S(4142) diagrams.  The slope of the dashed line is --0.476 with a rms of 0.40 \AA, which we have used as the error for the W(G) values.  The dotted lines are again set as $\pm$2 $\times$ rms.  Most of the stars lie inside the $\pm$2 $\times$ rms interval, but we see that the stars with large values of S(3839) and S(4142), i.e., the CN-strong stars, have generally lower values of W(G). Table \ref{Tabla_slopes} lists the information from the fits shown in Figures \ref{fig_panel_SMC_3839} and \ref{fig_panel_SMC_4142_wg}.

\begin{table}
\caption{Details of the fit on the figure \ref{fig_panel_SMC_3839} and \ref{fig_panel_SMC_4142_wg}}
\renewcommand{\TPTminimum}{\linewidth}
\centering 
\begin{adjustbox}{max width=8.5cm}
\setlength{\tabcolsep}{10pt} 
\renewcommand{\arraystretch}{1.3} 
\begin{tabular}{l c c c c} 
\hline 
\hline
\normalsize{} & \normalsize{\textbf{W(G) }} & \normalsize{\textbf{S(3839)} }   & \normalsize{\textbf{S(4142) }}  \\ [0.2ex] 
\hline 

Slope						& -0.476 & -0.123	    &  -0.073	&\\
Error in the slope			& 0.160  & 0.021		& 	0.012  &\\
Intercepts					& 10.18  & -0.128		&	-0.365  &\\
rms							& 0.40  & 0.05		&	0.03   &\\

\hline 

\end{tabular}
\end{adjustbox}
\label{Tabla_slopes} 
\end{table}

By using the fits to the CN-weak stars, we have calculated a parameter $\delta$  for each index. The value of $\delta$ is the vertical displacement of the observed index values with respect to the line-fit at the $V-V_{HB}$ of the star. Use of  $\delta$ values helps to minimize the effect on the band strengths of the different temperatures and surface gravities of the  observed stars \citep{norris1981cyanogen}.
 As for the original works in this subject field (e.g., \citet{norris1979cyanogen, norris1981abundance}) we show in Figure \ref{gen_hist} generalized histograms of the $\delta$ parameter for each index. In these generalized histograms  the sigma has been taken as the rms value about the fit to CN-weak stars. The adopted values of sigma (also given in Table \ref{Tabla_slopes}) are shown in the upper-right corner of each panel in the figure.   As is the case for many GGCs (e.g., \citet{smith1983cyanogen}), a bimodal distribution is quite clear in the $\delta$(S3839) histogram, while the $\delta$S(4142) histogram also shows some indication of bimodality.  Bimodality is not evident in the $\delta$W(G) histogram, though it is skewed to lower values.

In the upper panel of Fig.\ \ref{correlation_SMC} we have plotted $\delta$(WG) against $\delta$S(3839). 
A clear anti-correlation appears, which follows the behaviour seen for GGCs.  The correlation coefficient for the data points is  \textit{r} = --0.53, indicating for 16 degrees of freedom that there is a $< 5 \%$ probability that the observed anti-correlation arises by chance.  The observed data points, however, are effectively a single realization from a distribution governed by the errors in the $\delta$ values.  We have investigated the effect of this by conducting multiple trials in which each observed $\delta$ value is randomly perturbed to new value by using a gaussian distribution with mean zero and a standard deviation equal to the uncertainty in the $\delta$ value.  These uncertainties are taken as the rms values given in Table \ref{Tabla_slopes}.  We find that after 10,000 trials the mean value of \textit{r} is --0.41 with a standard deviation of 0.14: the observed \textit{r} is then a $<1\sigma$ deviant, and the mean \textit{r} value indicates a  $< 10 \%$ probability that the anti-correlation arises by chance.

The lower panel of Fig.\ \ref{correlation_SMC} shows, as expected, a good correlation between the $\delta$(S4142) and the $\delta$(S3839) values.  The value of  \textit{r} is 0.67 indicating a $< 1 \%$ probability that correlation arises by chance.  Following the same perturbation process, the mean value of \textit{r} for 10,000 trials is 0.52 with a standard deviation of 0.13; with this \textit{r} value, the probability of the correlation arising by chance is $< 5 \%$.  We conclude therefore that the anti-correlation and correlation shown in the panels of Fig.\ \ref{correlation_SMC}  are statistically significant.  The line strength differences are further illustrated in Fig. \ref{synth_example_chcn_overplotted} where we have overplotted the continuum-normalized spectra for two stars with similar colours and $V-V_{HB}$ magnitudes, and thus similar $T_{eff}$ and log $g$ values.  The stars are ID 1564 (CN-strong) and ID 1262 (CN-weak).  The spectra confirm that ID 1564 has notably stronger $\lambda$3883\AA \ CN-band strength but the difference at the G-band of CH is much less substantial.

Table \ref{table_disper_param} lists the mean values of the $\delta$ values for each index and for the CN-strong and CN-weak stars separately.  Also given are the standard deviations and the number of stars in each group.  These values reinforce our results: for example, the mean $\delta$S(3839) for the CN-strong and CN-weak stars is approximately 10.5$\times$ the combined standard deviation of the means, while for $\delta$S(4142) the difference in the means is $\sim$5.1$\times$  the combined standard deviation of the means.  Both of these are statistically significant.  For  $\delta W(G)$ however, while the difference is in the sense that the CN-strong stars have smaller values than for the CN-weak stars, the difference, at 2.5$\times$ the combined standard deviation of the means, is only marginally significant.

%
%

\begin{figure}
  \centering
  \includegraphics[width=8.5cm]{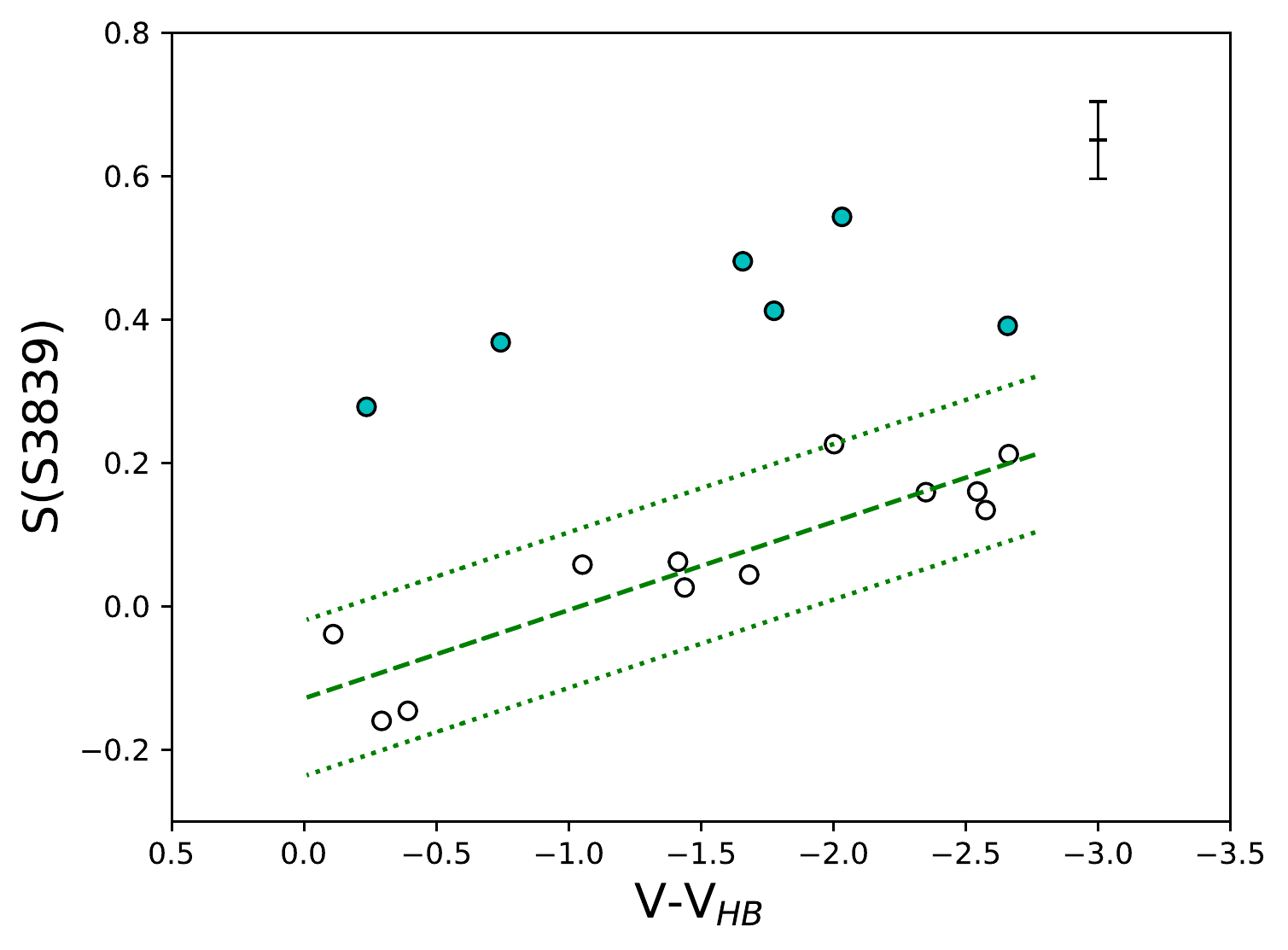}
  \caption[Dependence of the CN band-strength index S(3839) on $V-V_{HB}$ for RGB members of Kron~3]{Dependence of the CN band-strength index S(3839) on $V-V_{HB}$ for RGB members of Kron~3.  Filled symbols are used for CN-strong stars (defined by this diagram) and open symbols for CN-weak stars.  The green dashed line is the best fit to the index values for the CN-weak stars, and the dotted lines show $\pm$ 2 $\times$ rms of the fit.  The adopted $\pm$1$\sigma$ index error value is shown in the top right corner.}
\label{fig_panel_SMC_3839}
\end{figure}

\begin{figure}
  \centering
   \includegraphics[width=8.5cm]{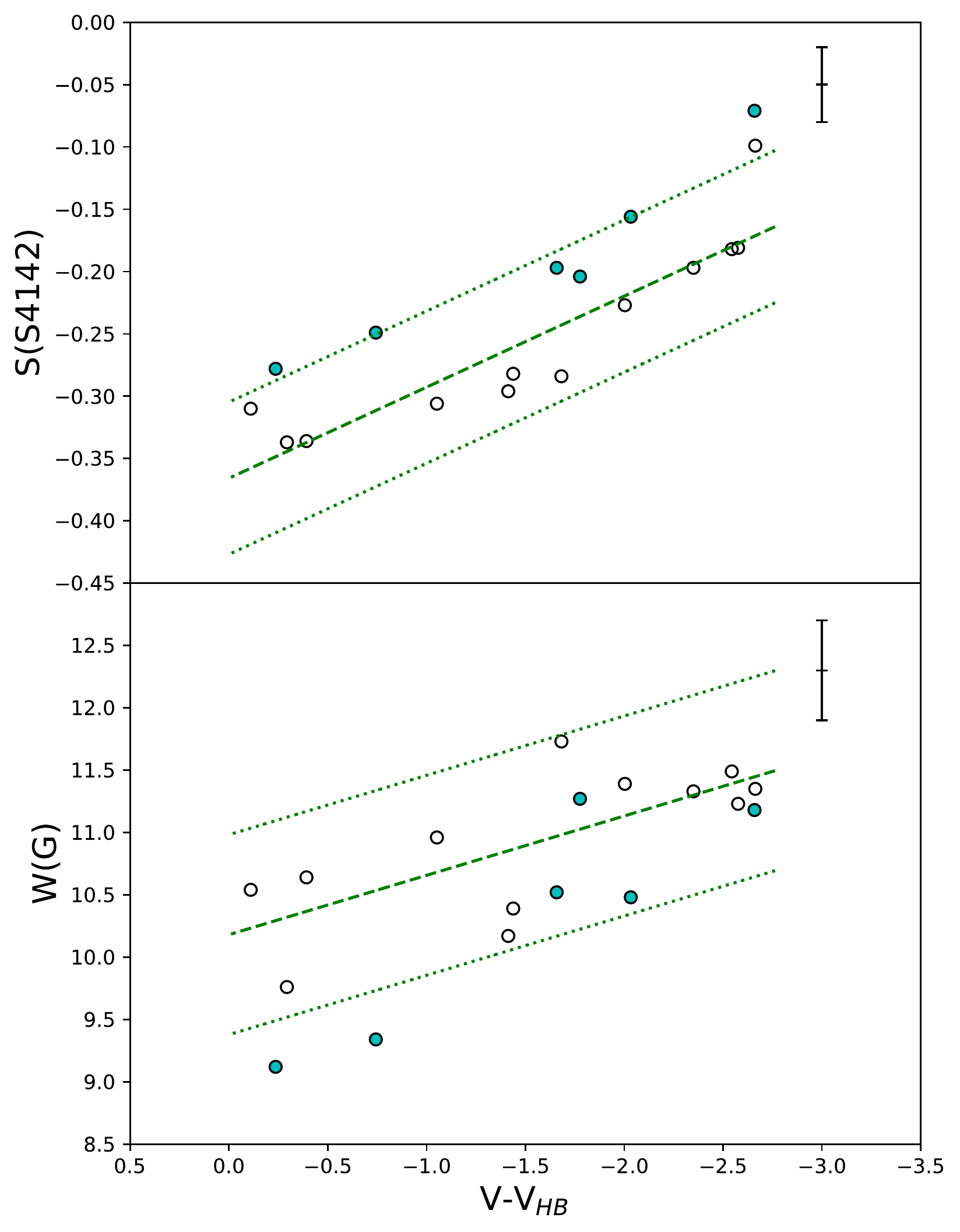}
  \caption[Upper panel: Dependence of CN band-strength index S(4142) on $V-V_{HB}$ for RGB members of Kron~3.]{Upper panel: the dependence of the CN band-strength index S(4142) on $V-V_{HB}$ for RGB members of Kron~3. Lower panel: the dependence of the CH band-strength index W(G) on $V-V_{HB}$ for RGB members of Kron~3. In both panels the symbol and line definitions are as for Figure  \ref{fig_panel_SMC_3839}.  Adopted $\pm$1$\sigma$ index error values are shown in the top right corner of each panel.}
\label{fig_panel_SMC_4142_wg}
\end{figure}


\begin{figure}
\begin{center}
\includegraphics[width=8.5cm]{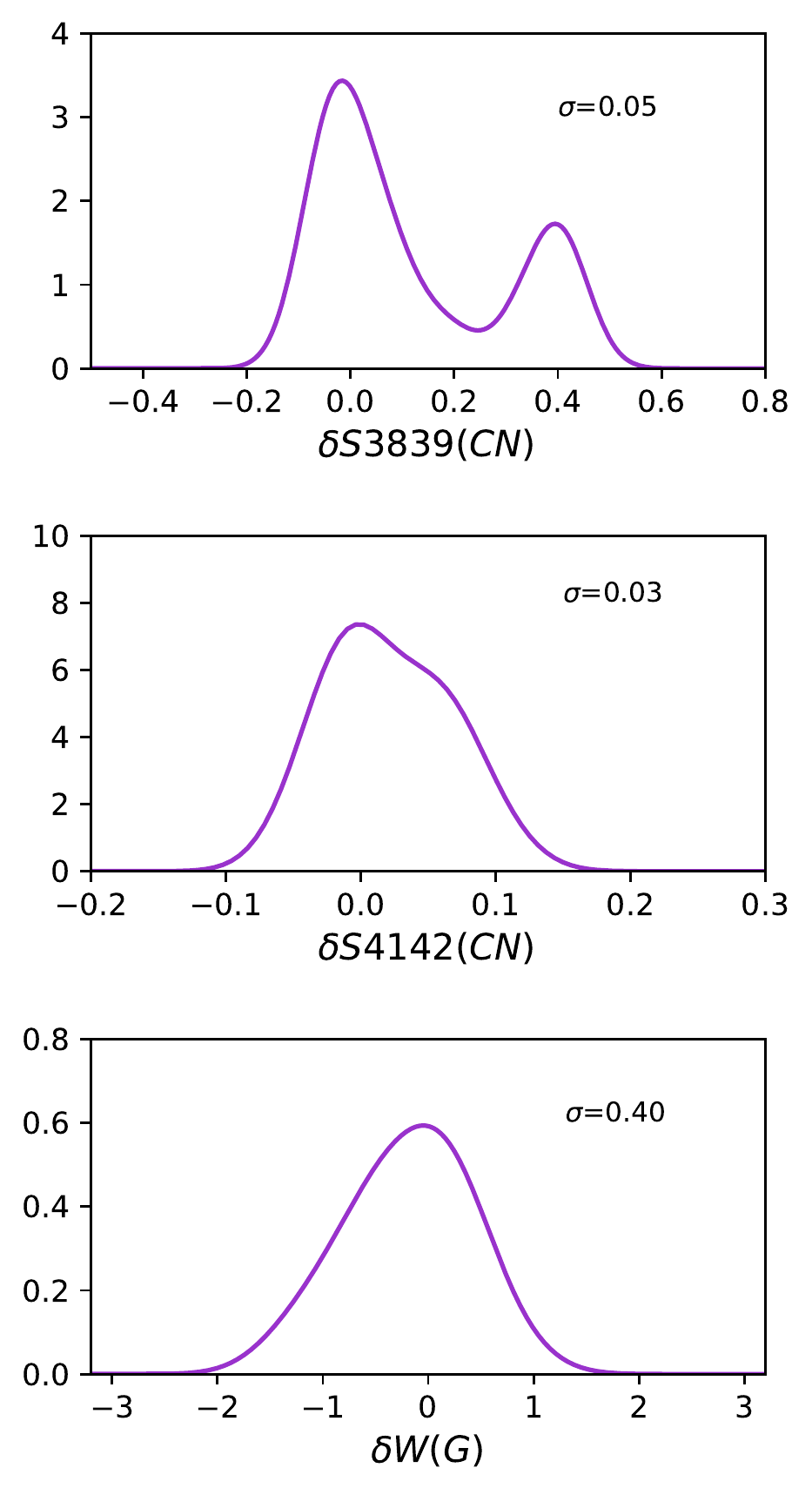}
\caption{Generalized histograms of $\delta S3839$ (top row), $\delta S4142$ (middle row) and $\delta W(G)$ (bottom row) for the Kron~3 cluster members.  The sigma ($\sigma$) of the gaussians used in generating the histograms are given in top-right of each panel.} 

\label{gen_hist}
\end{center}
\end{figure}

\begin{figure}
  \centering
  \includegraphics[width=8.5cm]{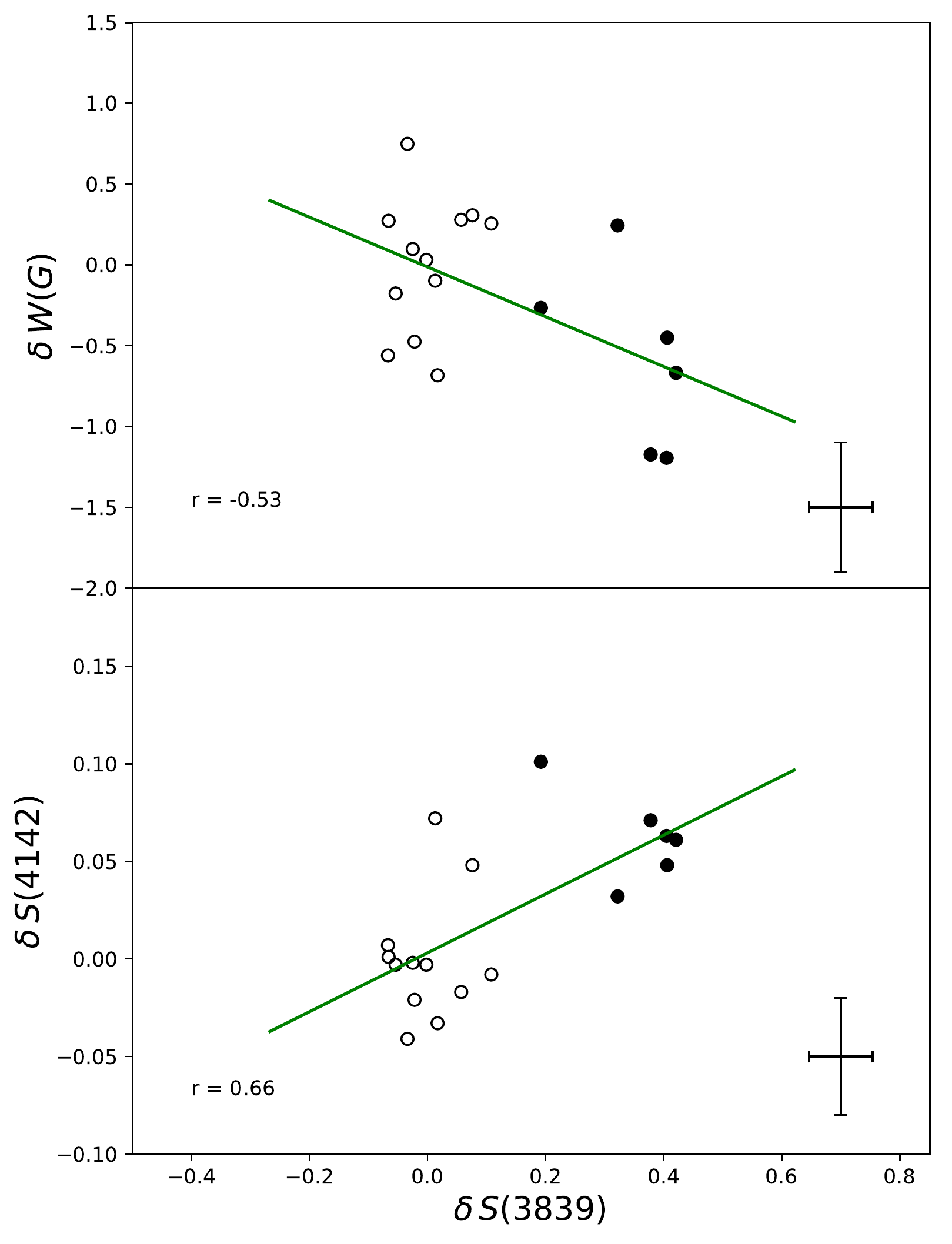}
  \caption[Dependence of $\delta W(G)$ on $\delta S(3839)$  for RGB stars of 
Kron~3]{Upper panel: Dependence of $\delta W(G)$ on $\delta S(3839)$  for RGB members of 
Kron~3. Lower panel: Dependence of $\delta S(4142)$ on $\delta S(3839)$  for the same stars. Filled and empty circles represent CN-strong and CN-weak stars as defined by Fig.\ \ref{fig_panel_SMC_3839}. The green line represents the best fit to the data points and the corresponding correlation coefficient \textit{r} is given in the panels. Error bars ($\pm1\sigma$) for each index are shown in the bottom right corners.}
\label{correlation_SMC}
\end{figure}

\begin{figure}
\begin{center}
\includegraphics[width=8.5cm]{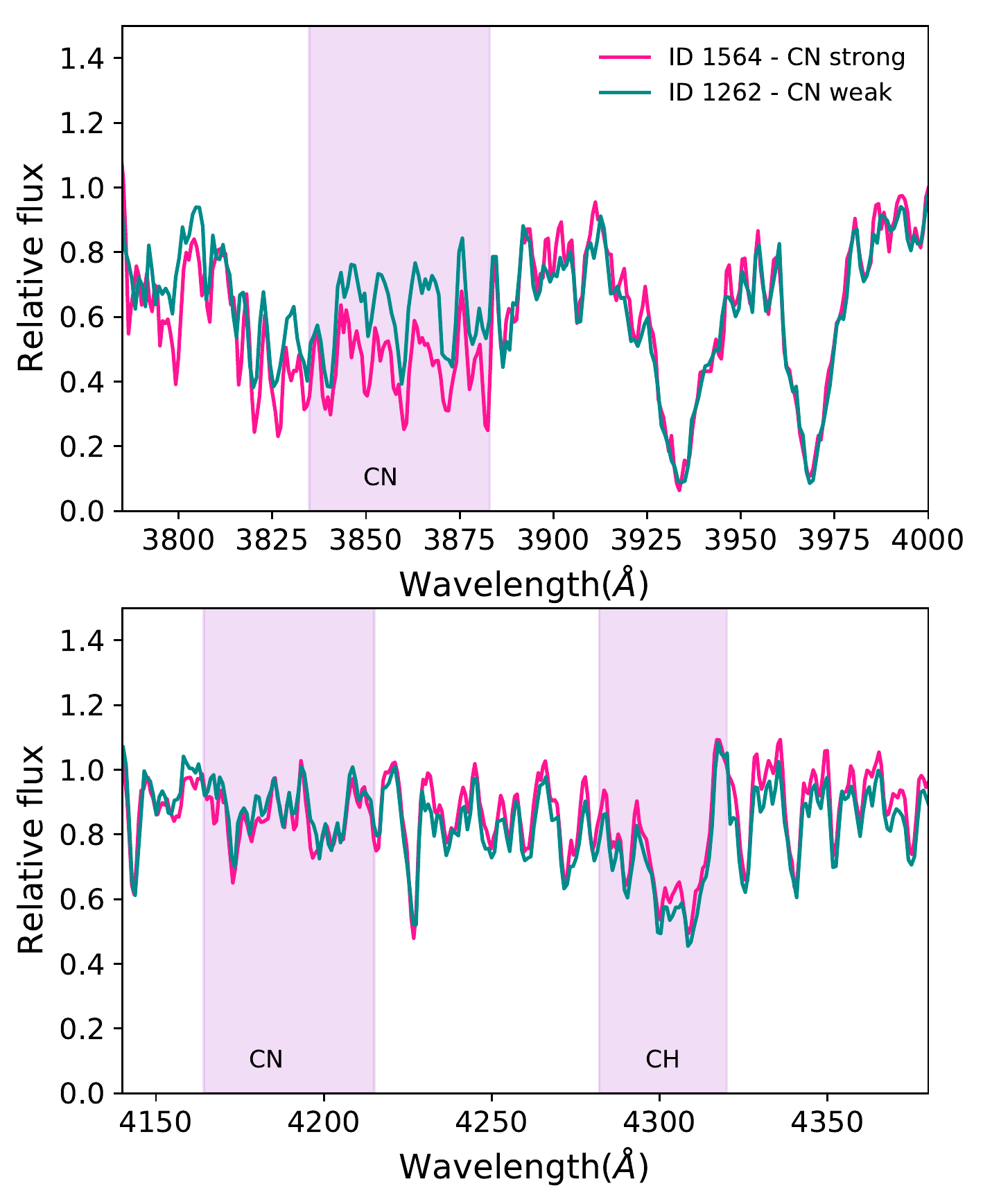}
\caption{Continuum normalized spectra of target stars ID 1564 (CN-strong, pink) and ID 1262 (CN-weak, turquoise). The location of the CH, at $\lambda \sim$ 4300 \AA \ and the CN, at $\lambda \sim$ 3883  \AA \  and $\lambda \sim$ 4215 \AA \, bands  are shown by the shaded regions. Star ID 1564 has S(3839) = 0.481 while for Star ID 1262 the value is 0.044 mag.  Similarly, the W(G) values are 10.52 and 11.73 \AA, respectively.}
\label{synth_example_chcn_overplotted}
\end{center}
\end{figure}


\begin{table}
\caption{Mean and standard deviation $\sigma_{\delta}$ of the $\delta$ indices for the two groups of Kron 3 stars.  The number of stars in each group is also listed.}
\resizebox{\linewidth}{!}
{\begin{threeparttable}
\renewcommand{\TPTminimum}{\linewidth}
\centering 
\begin{adjustbox}{max width=8.5cm}
\begin{tabular}{l c c c r} 
\hline 
\hline
\normalsize{} & $\quad$ \normalsize{Mean}$\quad$ & $\quad$ \normalsize{$\sigma_{\delta}$} $\quad$ & $\quad$ \normalsize{N of stars}$\quad$ &\\ [0.5ex] 
\hline 
\textbf{Kron~3} 									&  			& 				& 				&\\
$\delta W(G){\tiny _{weak}}$ 			&   0		&	0.40		&	12	&\\
$\delta W(G){\tiny _{strong}}$ 	  	&  -0.584	&	0.51		&	6	&\\
$\delta S(3839){\tiny _{weak}}$ 	  	&  0		&	0.05 		&	12 	&\\
$\delta S(3839){\tiny _{strong}}$ 		&	0.354	&	0.08		&	6	&\\
$\delta S(4142){\tiny _{weak}}$ 		&	0		&	0.03		&	12	&\\
$\delta S(4142){\tiny _{strong}}$ 		&	0.062	&	0.02		&	6	&\\
\hline 

\end{tabular}
\end{adjustbox}
\end{threeparttable}}
\label{table_disper_param} 
\end{table}

	\subsection{C, N abundances via spectrum synthesis }
 	\label{synth_smc}

The observed CN/CH band-strength anti-correlation in GGC red giants is driven by anti-correlated changes in the [N/Fe] and [C/Fe] abundances. In order to confirm  an analogous interpretation of Figs.\ \ref{fig_panel_SMC_3839}, \ref{fig_panel_SMC_4142_wg} and \ref{correlation_SMC} we have computed synthetic spectra for all  Kron 3 members. The spectral synthesis was performed in the same way as described in \citet{salgado2019investigation}.

In particular, the effective temperatures $T_{eff}$ were obtained using the reddening-corrected (B-V) colours and the $T_{eff}$-colour calibration of \citet{alonso1999effective}.  The Kron 3 reddening used was E(B-V) = 0.031 \citep{udalski1998optical}. The bolometric magnitudes were derived from the RGB V-band magnitudes and the bolometric corrections described in \citet{alonso1999effective}. The SMC distance modulus was taken as $(m - M)_0$ = 18.88 $\pm$ 0.1 mag  \citep {glatt2008age}.\footnote{\small{This value is consistent with more recent determinations of the SMC distance modulus (e.g., \citet{graczyk2020distance})}.}  We assumed $M_{*}$ = 1 solar mass for the RGB stars and  the surface gravities were calculated from the stellar parameters. The metallicity used in generating the synthetic spectra was that listed in Table \ref{cluster_parameter}, and the microturbulence velocity was set as $\xi_{t}$ = 2 km\,s$^{-1}$ for all stars.  The synthetic spectra were then smoothed to the resolution of the observed spectra. The derived effective temperatures and gravities are listed in Table \ref{tabla_total_abund_SMC}. 

In order to perform the synthetic spectrum calculations for [C/Fe] and [N/Fe], it is necessary to assume an [O/Fe] value for the stars.\footnote{\small{The abundance of oxygen can affect the measurement of carbon  as the amount of free C to form CH is governed by the abundance of CO.}}
We have chosen [O/Fe] = +0.2 for both the CN-weak and  CN-strong stars. The consequences of this assumption were investigated by repeating the calculations using [O/Fe] = --0.2 for the CN-strong stars.  This showed that the changes in [O/Fe] does not have a strong influence on the derived [C/Fe] and [N/Fe] values.  Specifically, using [O/Fe] = +0.2 for the CN-strong stars resulted in [C/Fe] values approximately 0.08 dex smaller than for the [O/Fe] = --0.2 case, while the derived [N/Fe] values are $\sim$0.05 dex larger with [O/Fe] = +0.2 compared to the [O/Fe] = --0.2 case.

\begin{figure*}
\begin{center}
\includegraphics[width=12cm]{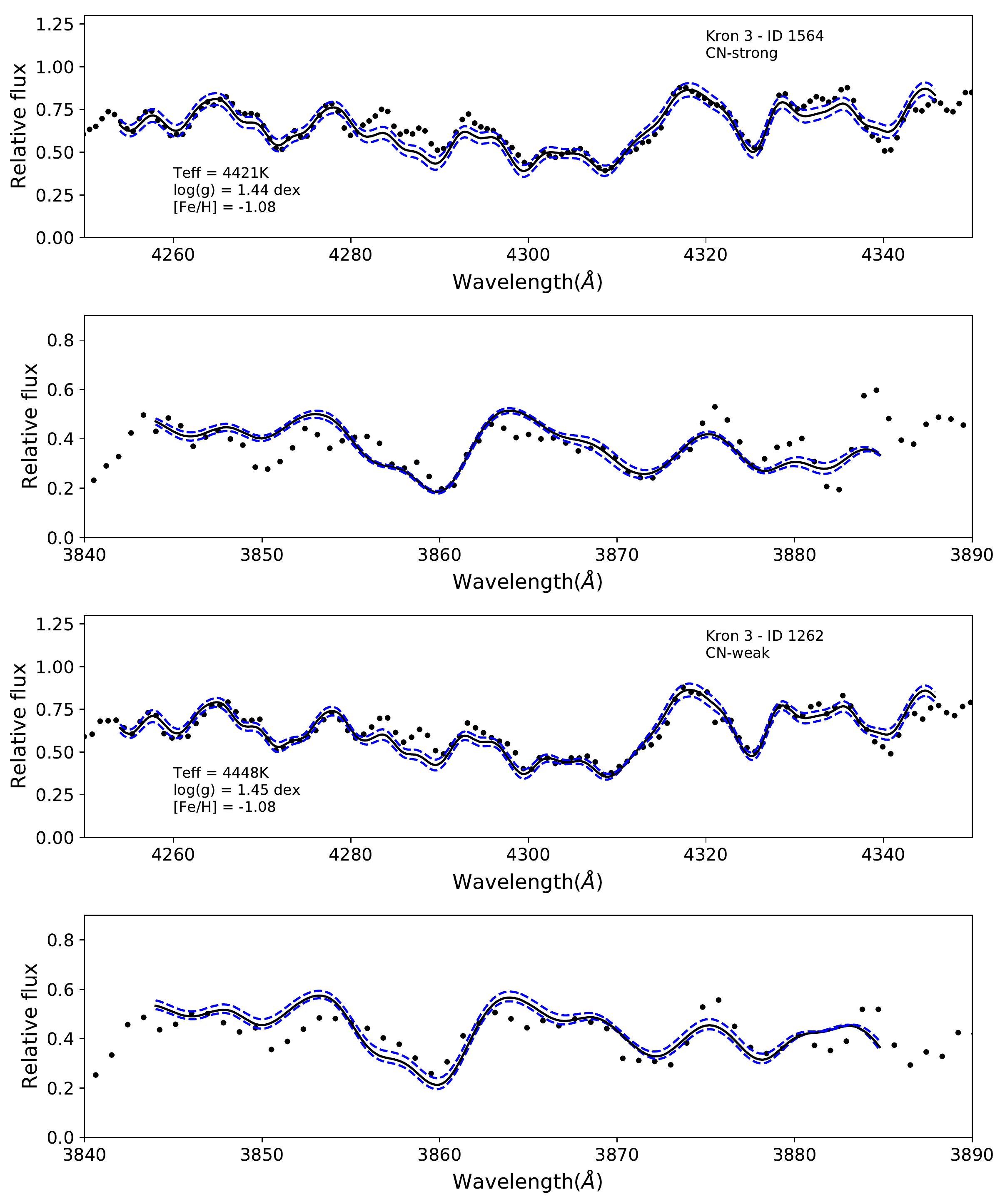}
\caption{Spectrum synthesis of CH (first and third panels) and CN (second and fourth panels) features for stars 
ID~1564 (CN-strong) and ID~1262 (CN-weak). In all panels the continuum normalized spectrum is represented by solid dots. First panel: the solid line represents the best abundance fit  which has [C/Fe] = --0.90, assuming [O/Fe] = +0.2. The dashed blue lines show [C/Fe] values $\pm$0.15 dex about the central value. Second panel: the solid line represents the best abundance fit with  [N/Fe] = +0.15 and [C/Fe] = --0.9, assuming [O/Fe] = +0.2. The dashed blue lines show [N/Fe] values $\pm$0.3 dex from the best fit. Third panel: the solid line represents the best abundance fit  which has [C/Fe] = --0.70 assuming [O/Fe] = +0.2. The dashed blue lines again show [C/Fe] values $\pm$0.15 dex about the central value. Fourth panel: the solid line represents the best abundance fit with  [N/Fe] = --0.55 and [C/Fe] = --0.70, assuming [O/Fe] = +0.2. The dashed blue lines show [N/Fe] values $\pm$0.3 dex from the best fit. The observed spectra are identical to those shown in Fig.\ \ref{synth_example_chcn_overplotted}. }

\label{synth_example}
\end{center}
\end{figure*}

As regards the fits to individual stars, we followed the same procedure as that in \citet{salgado2019investigation} for RGB stars in the Sculptor dwarf spheroidal galaxy.  First, the carbon abundance, for the adopted oxygen abundance, was determined via minimization of the residuals between the synthetic and observed spectra in the region of the G-band ($\lambda\, \approx$ 4300 \AA). Then, using the region 3840 -- 3885 \AA\/, the nitrogen abundance was determined using as input the value of oxygen abundance and the previously determined carbon abundance.\footnote{\small{Because of the difficulty in defining the actual continuum level at blue and near-uv wavelengths in the observed spectra of these cool red giants, we do not claim that the [C/Fe] and [N/Fe] values determined are valid in an absolute sense, but we do contend that they are valid in a differential sense.}} The process is illustrated in Figure \ref{synth_example} for the CN-strong star (ID~1564) and the CN-weak star (ID~1262). In the figure the best abundance fit is represented by solid black line: the abundances for star ID~1564 are [C/Fe] = --0.9 $\pm$ 0.17  and [N/Fe] =  0.15 $\pm$ 0.33, while they are [C/Fe] = --0.7 $\pm$ 0.18 and [N/Fe] =  --0.55 $\pm$ 0.33 for star ID~1262. These stars have similar temperatures: 4421 K for ID 1564 and 4450 K for ID 1262 (Table \ref{tabla_total_abund_SMC}). 

The errors,  $\sigma_{total}$, in the derived [C/Fe] and [N/Fe] abundances  were calculated  in the same way as discussed in \citet{salgado2016scl} and \citet{salgado2019investigation}, i.e., as a combination of errors from the uncertainties in the stellar parameters ($\sigma_{SP}$) with additional sources of error ($\sigma_{fit}$) that include the effects of the signal-to-noise of the spectrum and the fitting uncertainty. We estimated $\sigma_{SP}$ by repeating the analysis for a representative star from the sample, varying the atmospheric parameters by $\Delta T_{eff} = \pm$ 200 K, $\Delta$log$g = \pm$ 0.2, $\Delta [M/H]= \pm$ 0.4 and 
$\Delta \xi = \pm$ 0.2 km s$^{-1}$. The quantity $\sigma_{fit}$ comes from the comparison of the observed spectrum
with model spectra with different C and N abundances.  As the derived carbon abundance is dependent on the assumed oxygen abundance, for $\sigma_{fit\,C}$ we have added in quadrature a further $\sigma_{fit} = \pm 0.1$ to allow for the uncertainty in adopted oxygen abundances. In the same manner, since the determination of the
N abundance from the CN features depends of the derived carbon abundance, we have added in quadrature to $\sigma_{total}$ for nitrogen the value of $\sigma_{total}$ for carbon. Typical values
of $\sigma_{total}$ are $\sim$0.17 dex for [C/Fe] and $\sim$0.33 dex for [N/Fe]. The values of $\sigma_{total}$ are listed in the Table \ref{tabla_total_abund_SMC} for each star and abundance. 

\begin{figure}
\begin{center}
\includegraphics[width=8.5cm]{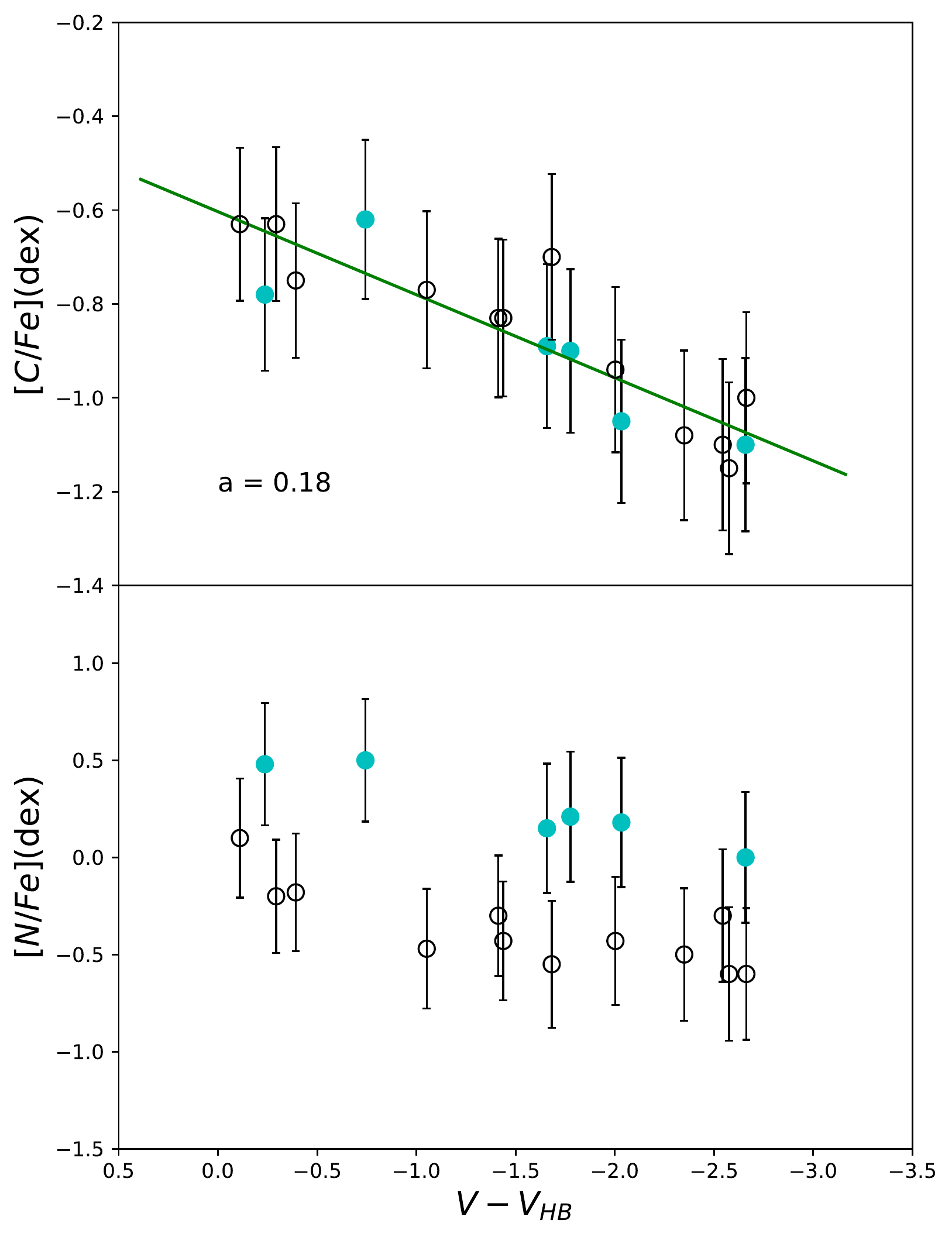}
 \caption{Upper panel: dependence of [C/Fe] on V magnitude for RGB members of the Kron~3. The green line represents the best fit and the corresponding slope value `a' is given on the figure. Lower panel: dependence of [N/Fe] on V magnitude for RGB members of the Kron~3.  In both panels filled and empty circles represent CN-strong and CN-weak stars. }
\label{evolutionary_mixing}
\end{center}
\end{figure}

In Fig.\ \ref{evolutionary_mixing} we have plotted the [C/Fe] and [N/Fe] abundances of all Kron 3 members as function of their $V-V_{HB}$ magnitude. It is evident from the upper panel of Fig. \ref{evolutionary_mixing} that the carbon abundance decreases with increasing luminosity, a signature of evolutionary mixing on the red giant branch.  Such evolutionary mixing has been observed in both field and globular cluster red giants (see, e.g., \citet{placco2014carbon} and references therein).  The depletion in carbon reflects CNO-cycle processing in the envelope of the star in which C is converted to N.  A decline in carbon should therefore be matched by a corresponding increase in nitrogen.  However, the lower panel of Fig.\ \ref{evolutionary_mixing} shows that despite the decrease in C with increasing luminosity, no corresponding increase in N is evident.  In particular, assuming initially scaled-solar abundances, and that the depleted C is entirely converted to N, then the decrease in [C/Fe] of $\sim$0.4 dex between the least and most luminous stars in the observed sample should result in an increase in [N/Fe] of $\sim$0.6 dex over the same $V-V_{HB}$ range.  Such an increase is not apparent in the lower panel of Fig. \ref{evolutionary_mixing} where there is no obvious trend between [N/Fe] and luminosity. It is possible that the expected trend is disguised by systematic effects in the N-abundance fits as a function of $V-V_{HB}$.  Such systematic effects, however, should not affect N-abundance differences between stars at similar $V-V_{HB}$ values.

\begin{figure}
\begin{center}
\includegraphics[width=8.5cm]{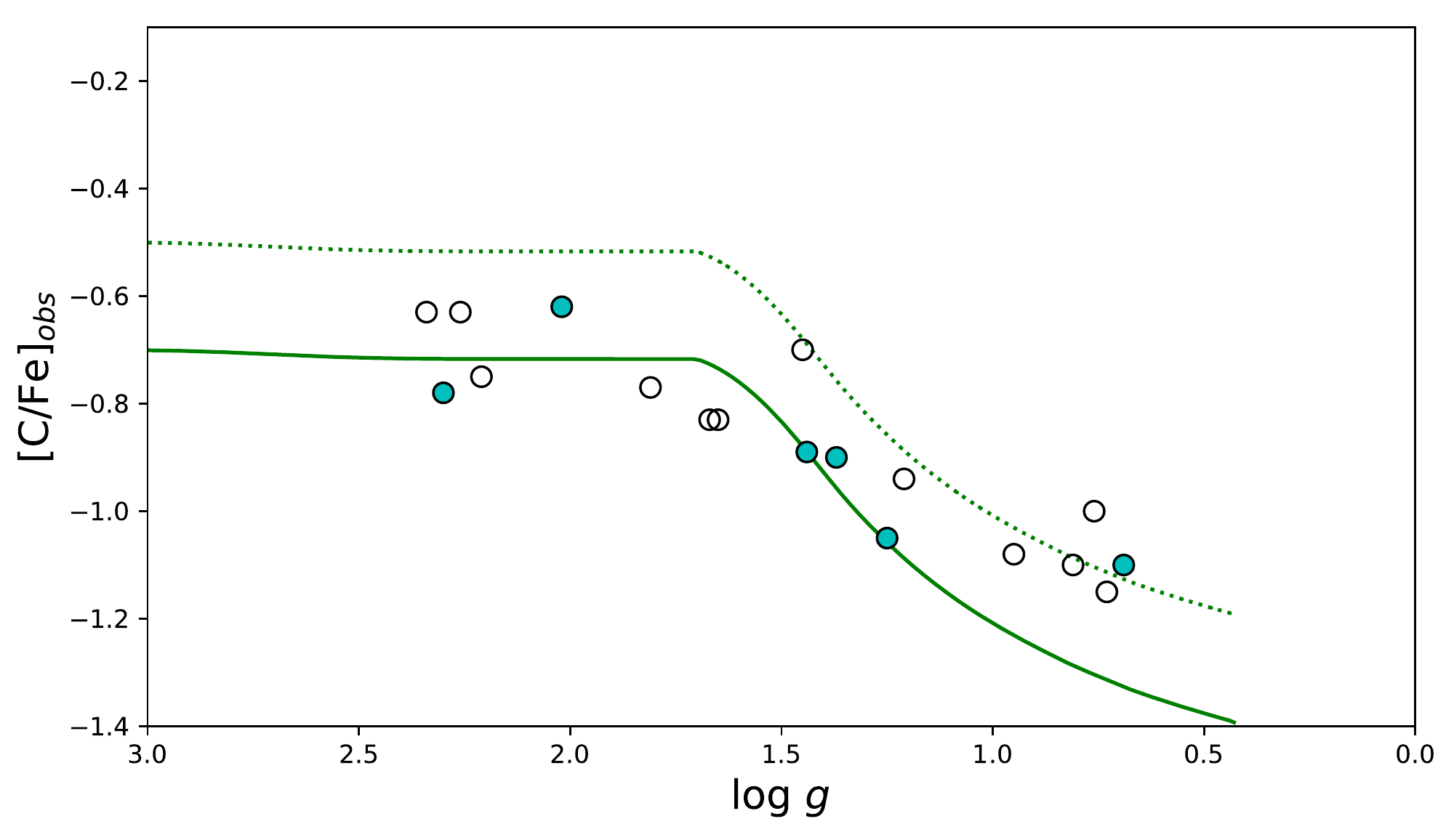}
\caption{[C/Fe] as function of log $g$ for the Kron 3 members. The dotted curve represents the \citet{placco2014carbon} model with [C/Fe] = --0.5 and [Fe/H] = --1.3. The solid line is the same dotted curve but shifted downwards by 0.2 dex. Filled and empty circles represent CN-strong and CN-weak stars.}
\label{placco_corr}
\end{center}
\end{figure}

Figure \ref{placco_corr} shows the observed carbon abundances as a function of the surface gravities (log $g$) of the stars.  Also shown on the figure as a dotted line is an evolutionary mixing C-depletion curve from \citet{placco2014carbon}(Fig. 6) for models with initial [C/Fe] = -0.5 and [Fe/H] = -1.3 dex.  The shape of the curve is similar to that of the observations, and in particular indicates that C-depletion does not occur for log $g$ $>$ 2.0 dex.  The solid curve is the same as the dashed line but with a downward shift of 0.2 dex applied\footnote{\small{We again note that absolute values of [C/Fe] in the observed stars are not well-established, but that the size of the variation with log $g$ is likely unaffected by any systematic uncertainty.}}, and it provides a reasonable fit to the observations except for the stars with log $g$ $<$ 1, where the model predicts larger depletions than shown by the observed stars.  Consequently, we have applied corrections for the effects of evolutionary mixing to the observed C abundances as discussed in \citet{placco2014carbon} using the web-based tool available at https:$//$vplacco.pythonanywhere.com, noting that the correction for the lowest log $g$ stars may be too large.  The corrections range from approximately zero for the highest gravity stars to $\sim$0.6 dex at the lowest gravities.

Use of the evolutionary mixing corrected carbon abundances necessarily implies that the observed nitrogen abundances should be decreased to compensate for the increased carbon abundances.  We assume that the depleted carbon has all been converted to nitrogen, and use the observed [C/Fe] values and the evolutionary mixing abundance corrections together to calculate the corresponding amount of nitrogen produced.  That value, under the assumption of an initial scaled-solar nitrogen abundance, allows the calculation of a N abundance correction, which is then applied to the observed [N/Fe] values.  The corrections, which reduce the observed N abundances, range from effectively zero for the stars with log $g$ $>$ 2 to $\sim$0.35 dex for the lower gravity stars.  The observed and evolutionary mixing corrected [C/Fe] and [N/Fe] values for the Kron 3 members are given in Table \ref{tabla_total_abund_SMC}.

\begin{figure}
\begin{center}
\includegraphics[width=8.5cm]{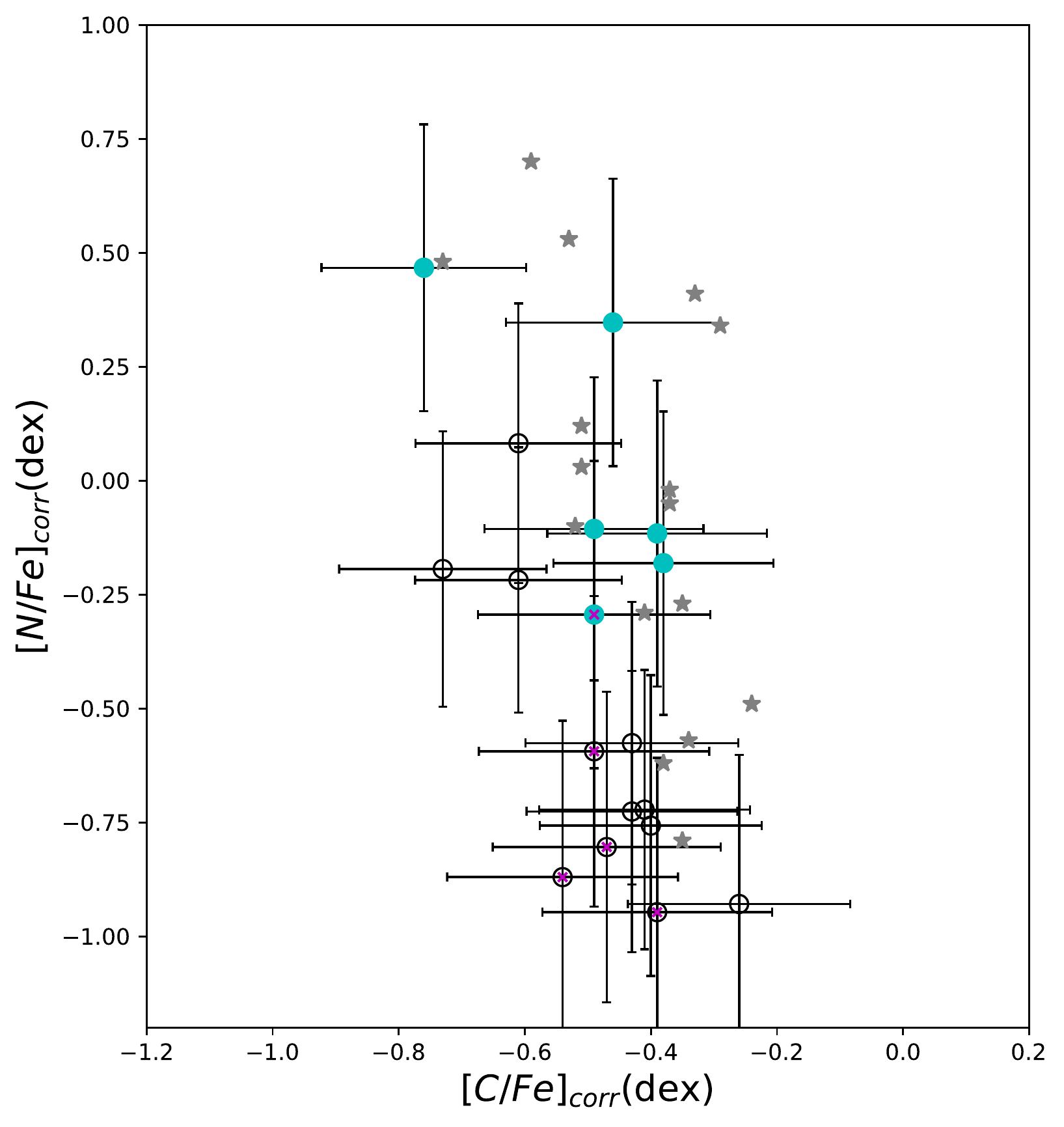}
 \caption{The relation between the [C/Fe] and [N/Fe] values after correction for evolutionary mixing for our sample of Kron~3 stars. 
 Filled and empty circles represent CN-strong and CN-weak stars, respectively, while the magenta coloured symbols represent the stars with log $g$ $<$1. Individual uncertainties are shown for each star.  In addition, the grey asterisk symbols correspond to 
 the [C/Fe] and [N/Fe] abundances for the independent set of Kron 3 stars reported in \citet{hollyhead2018kron}. 
   }
\label{abun_C_N_SMC}
\end{center}
\end{figure}

Our results for the carbon and nitrogen abundances corrected for evolutionary mixing effects are shown in Fig. \ref{abun_C_N_SMC}.  Also shown in the figure are the carbon and nitrogen abundances for the Kron 3 members studied by   \citet{hollyhead2018kron}.  The \citet{hollyhead2018kron} stars lie at fainter magnitudes in the cluster CMD than our sample so that there are no stars in common.  Further, as a consequence of the fainter magnitudes, there is no need to apply any evolutionary mixing corrections to the \citet{hollyhead2018kron} observed abundances.  It is clear from Fig. \ref{abun_C_N_SMC} that both data sets are in excellent agreement, which supports the application of the evolutionary mixing corrections to our observed abundances.  Specifically, the mean [C/Fe] abundance for our sample is only 0.065 dex lower than that for the \citet{hollyhead2018kron}  sample, while for [N/Fe], our mean abundance is 0.36 dex lower. Given the abundance uncertainties, and the independent analysis approaches, this agreement is reassuring.  Further, the observed range in [C/Fe] in both data sets is essentially identical at $\sim$0.5 dex.  This is also the case also for [N/Fe] where both data sets exhibit a range of $\sim$1.4 dex. \citet{hollyhead2018kron} do not claim that a C/N anti-correlation is present in their data, arguing that it may be hidden as the uncertainties in their C abundances are comparable to the C abundance range.  For our sample, however, there is an indication of C/N anti-correlation: the correlation coefficient is --0.57 (corresponding to a $< 2\%$ chance of the anti-correlation arriving by chance), although conducting multiple trials, assuming $\sigma$[C/Fe] = 0.17 and $\sigma$[N/Fe] = 0.33 (see Table \ref{tabla_total_abund_SMC}), yields a mean correlation coefficient of --0.28 ($\sigma$ = 0.20), a value that indicates there is a $>$10\% chance that the anti-correlation arises by chance.  We therefore concur with \citet{hollyhead2018kron} that establishing the reality of any C/N anti-correlation requires improved abundance ratio uncertainties.

In summary, the range in both C and N abundances seen in both the \citet{hollyhead2018kron} and our sample are directly comparable to the C and N ranges seen in GGCs.  As suggested by \citet{hollyhead2018kron}, this result is strong evidence for the existence of multiple populations in the $\sim$6.5 Gyr old SMC star cluster Kron 3 that are analogous to those seen in the ancient GGCs.  Confirmation of this assertion requires investigation of the Na abundances in our Kron 3 stars, as in the multiple populations hypothesis the N-rich stars should also be Na-rich. 

%

 	\subsection{NaD lines strengths and Sodium abundances}
	\label{sodium_smc} 

One of the defining characteristics of the GGC light element abundance variations is the O/Na anti-correlation \citep[e.g.,][]{carretta2009anticorrelation, carretta2016spectroscopic}. Since the oxygen depletion is accompanied by nitrogen enhancement, investigating a Na/N correlation is equivalent to seeking a Na/O anti-correlation.
As the variation in CN-bands is not sufficient in itself to claim the presence of abundance variations in stars with luminosities as for our sample (because of the effects of evolutionary mixing on RGB stars), the existence of a correlation between CN-band strengths and NaD line strengths, or more significantly, between N and Na abundances, is crucial to validate a multiple populations interpretation. 

In the same manner as was presented in \citet{salgado2019investigation} we have measured the strengths of the NaD lines and determined sodium abundances for our targets. We note that the Kron~3 radial velocity is sufficient to allow a clear separation of the stellar NaD lines from any interstellar component, which, in any case is likely to be small given the low reddening for the cluster.  Any differential reddening across the face of the cluster is likely even smaller.  Kron~3 is also well away from any active
star-forming region in the SMC \citep{glatt2008age}.
Because of smaller field of view of GMOS-S compared to FORS2, not all of the cluster members have NaD spectra. We use the CN-band strength as an indicator of the N abundance and seek evidence for a positive correlation between Na and N, first using band and line strengths, and second using abundances determined from synthetic spectrum fitting. 

\begin{figure}
\begin{center}
\includegraphics[width=8.5cm]{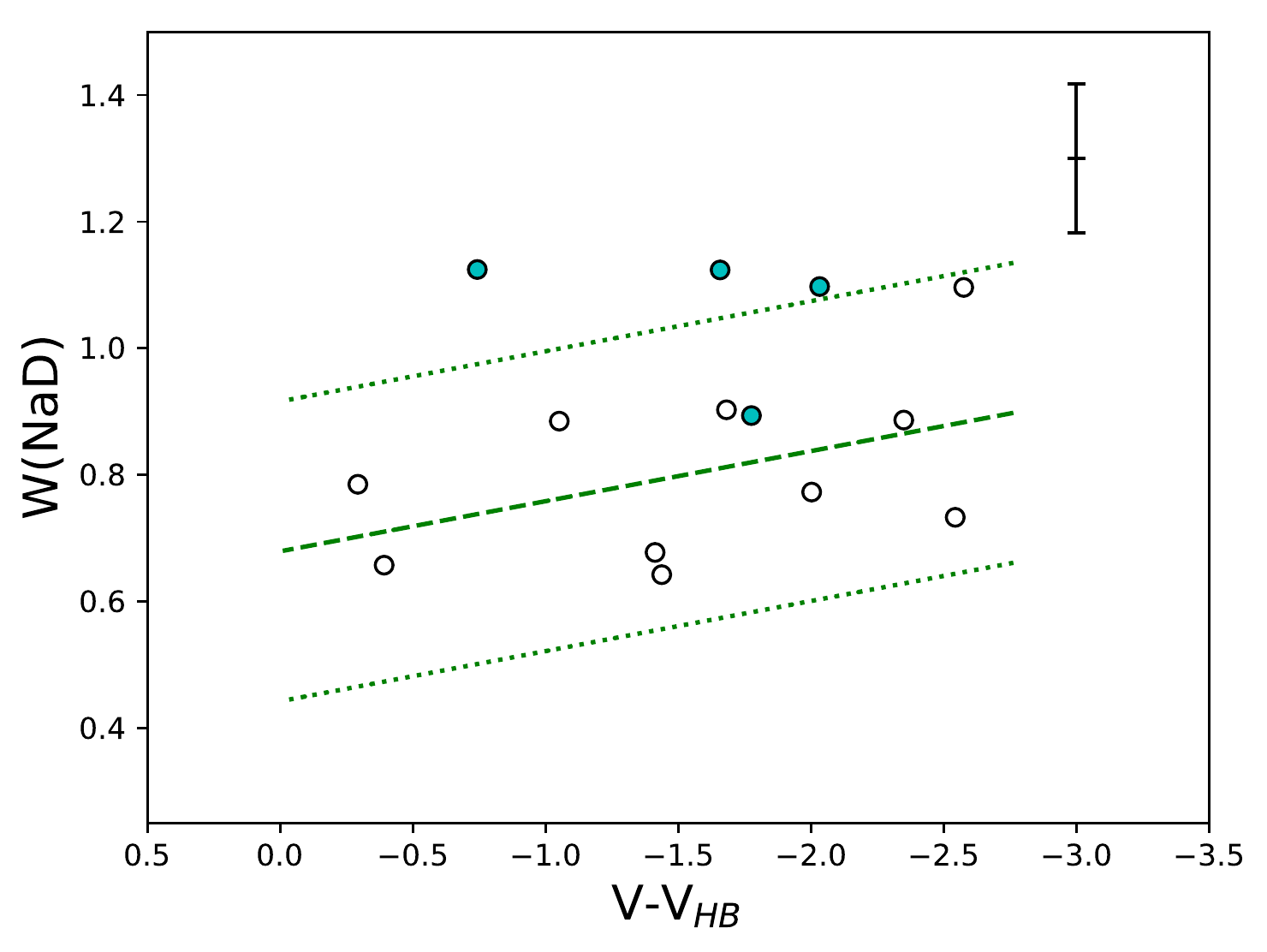}
\caption[Dependence of the line-strength index W(NaD) on $V-V_{HB}$ for RGB members of Kron 3.]{Dependence of the line-strength index W(NaD) on $V-V_{HB}$ for RGB members of Kron 3. CN-strong and CN-weak stars are represented by filled and open symbols, respectively. The green dashed line represents the best fit to the CN-weak stars and the dotted lines show $\pm$ 2 $\times$ rms of the fit. A typical error bar ($\pm$ 1 $\sigma$) is shown shown in the top right corner.}

\label{wnad_smc}
\end{center}
\end{figure}

Using the method presented in section \ref{strengths3} we have determined the equivalent width W(NaD) of the NaD lines at 
$\sim$5889 \AA \ and $\sim$5895 \AA.
The results are shown in Fig.\ \ref{wnad_smc}. As for the band strength measurements discussed in section \ref{results_CH_CN}, the dashed line in Fig. \ref{wnad_smc} is a least squares fit to the W(NaD) values for the CN-weak stars.  The slope is --0.079 \AA$/$mag and the rms about the fit is 0.118 \AA , which we take as an estimate of the error in the W(NaD) values.  The dotted lines are offset from the dashed line by $\pm$2$\times$rms.  A slight trend for W(NaD) to increase with increasing luminosity is evident.  This is reminiscent of the similar trends seen in the equivalent plots for the GGCs M55, NGC~6752 and NGC~288 discussed in  \citet{salgado2019investigation}. 

As for the band-strength indices, we allow for the change in W(NaD) with increasing luminosity by calculating for each star the value of $\delta$W(NaD), the vertical offset from the fitted line at the $V-V_{HB}$ of the star.  We find for the 4 CN-strong stars that the mean $\delta$ is 0.26 \AA\  with a standard error of the mean of 0.07 \AA.  For the 10 CN-weak stars the mean delta is 0.00 \AA\  (by construction) with a standard error of the mean of 0.04 \AA.  The difference in the means corresponds to 3.2 $\times$ the combined $\sigma$, indicating that the CN-strong stars do have stronger Na D lines compared to CN-weak stars of similar luminosity, similar to what is seen the GGCs (see, e.g., Fig. 17 of \citet{salgado2019investigation}).  The line strength difference is illustrated in Fig. \ref{synth_example_na_overplotted} where we have overplotted the spectra of two stars with similar $V-V_{HB}$ values: ID 1755 (CN-strong) and ID 1518 (CN-weak).  The difference in Na D line strengths is evident.

\begin{figure}
  \centering
  \includegraphics[width=8.5cm]{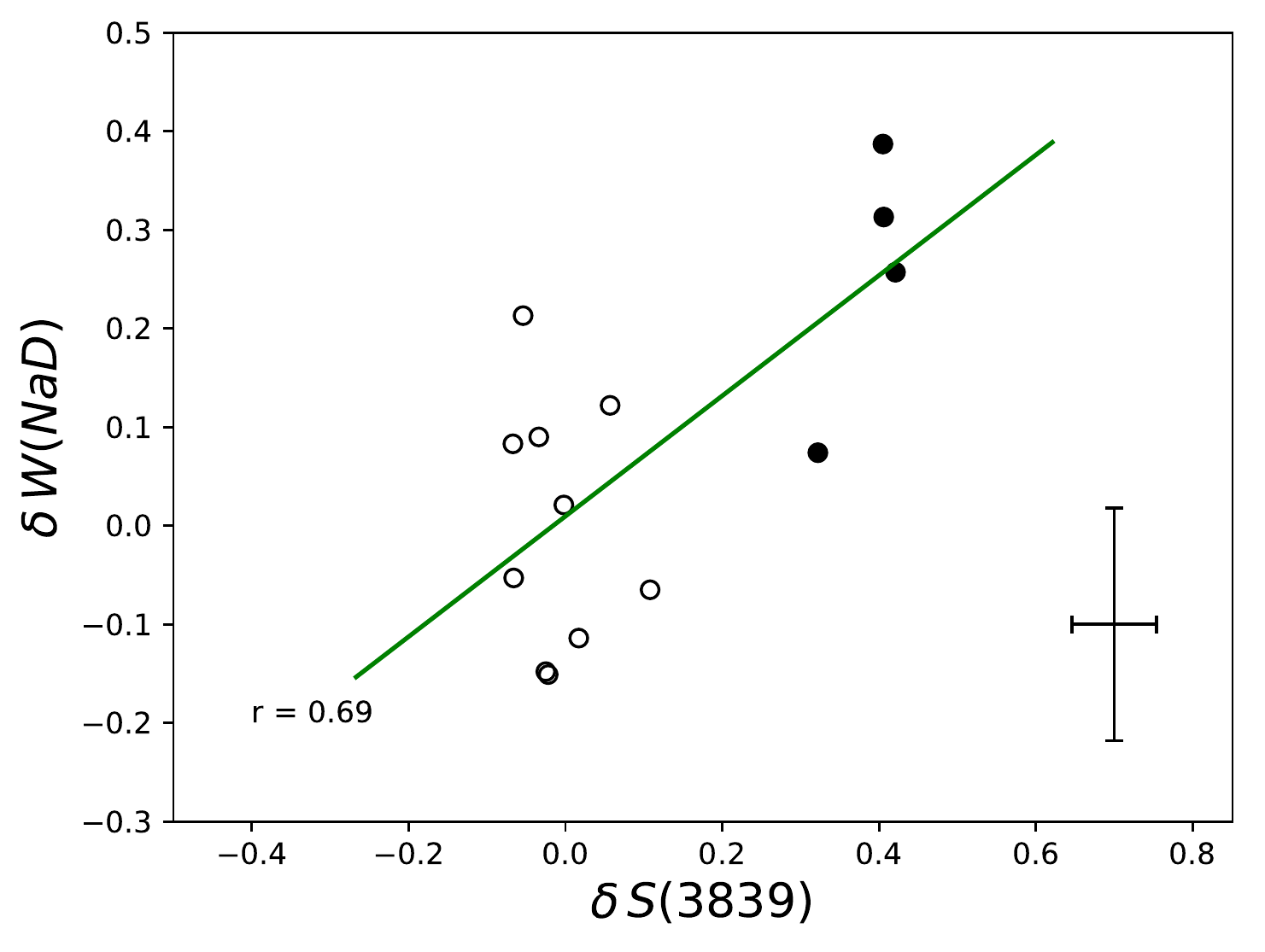}
  \caption[Dependence of $\delta$W(G) on $\delta$S(3839)  for RGB stars of 
Kron~3]{Dependence of $\delta$W(NaD) on $\delta$S(3839) for RGB members of 
Kron~3. Filled and empty circles represent CN-strong and CN-weak stars, as defined by Fig.\ \ref{fig_panel_SMC_3839}. The green line represents a least-squares fit. The correlation coefficient \textit{r} is given on the figure. Error bars are shown in the bottom right corner.}
\label{correlation_SMC_NAD}
\end{figure}

\begin{figure*}
\begin{center}
\includegraphics[width=12cm]{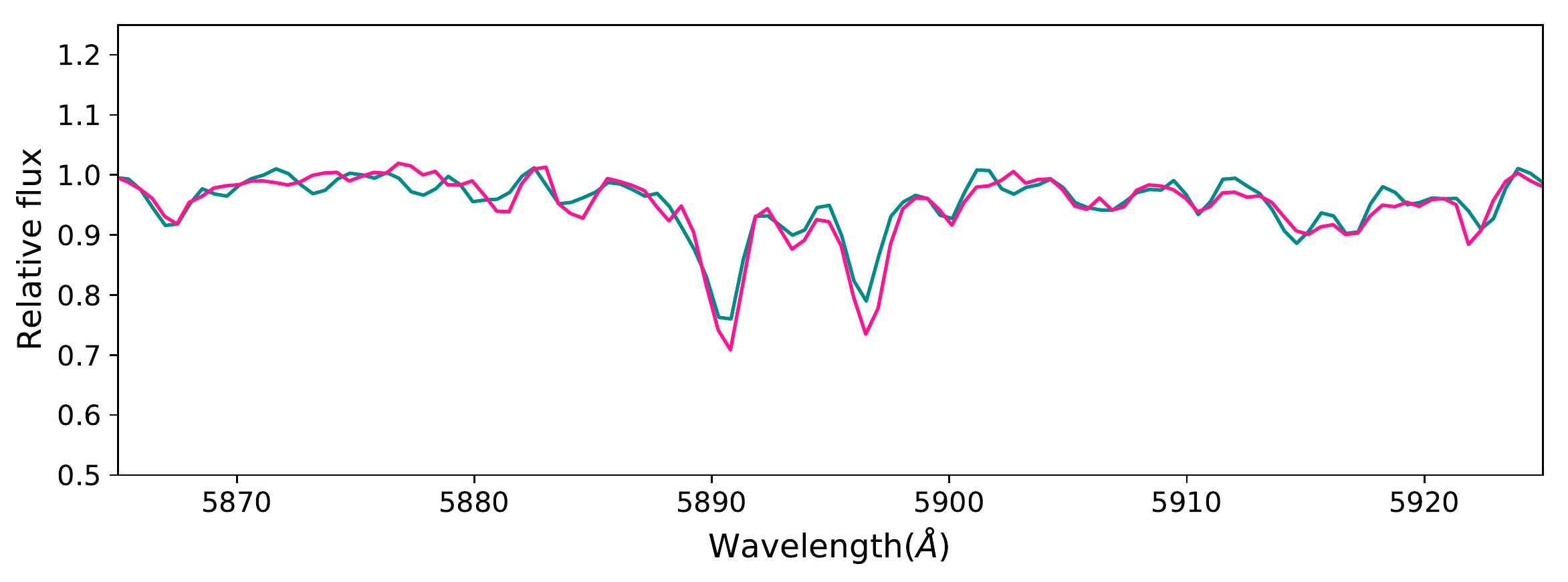}
\caption{Continuum normalized spectra of target stars ID 1755 (CN-strong, pink) and ID 1518 (CN-weak, turquoise) in the vicinity of the sodium D-lines at $\lambda$5889 and 5895 \AA . These stars have W(NaD) values of 1.097 and 0.772 \AA \  respectively}

\label{synth_example_na_overplotted}
\end{center}
\end{figure*}

In Fig. \ref{correlation_SMC_NAD} we have plotted $\delta$W(NaD) against $\delta$S(3839). The errors for each $\delta$ are, as before, those from the rms of fits to the CN-weak stars: 0.054 mag for $\delta$S(3839) and 0.118 \AA \ for $\delta$W(NaD).  They  are plotted in the bottom-right corner of the figure. The correlation coefficient for the points shown in Fig.\ \ref{correlation_SMC_NAD} is  \textit{r} = 0.69, which corresponds, for 12 degrees of freedom, to a $<$ 1\% probability that the observed correlation arises by chance.  Following the same procedure as explained in the Section \ref{results_CH_CN}, we have investigated the influence of the errors on the observed correlation by again conducting multiple trials in which each observed $\delta$ value is randomly perturbed to new value by using a gaussian distribution with mean zero and a standard deviation equal to the uncertainty in the $\delta$ value.  After 10,000 trials we find that the mean correlation coefficient is 0.55, with a standard deviation of 0.14; for this value of \textit{r} the probability that the correlation arises by chance remains $<$ 5\%.  We conclude that there is real correlation between CN-band strengths and NaD line strengths in our Kron 3 sample.

\begin{figure*}
\begin{center}
\includegraphics[width=12cm]{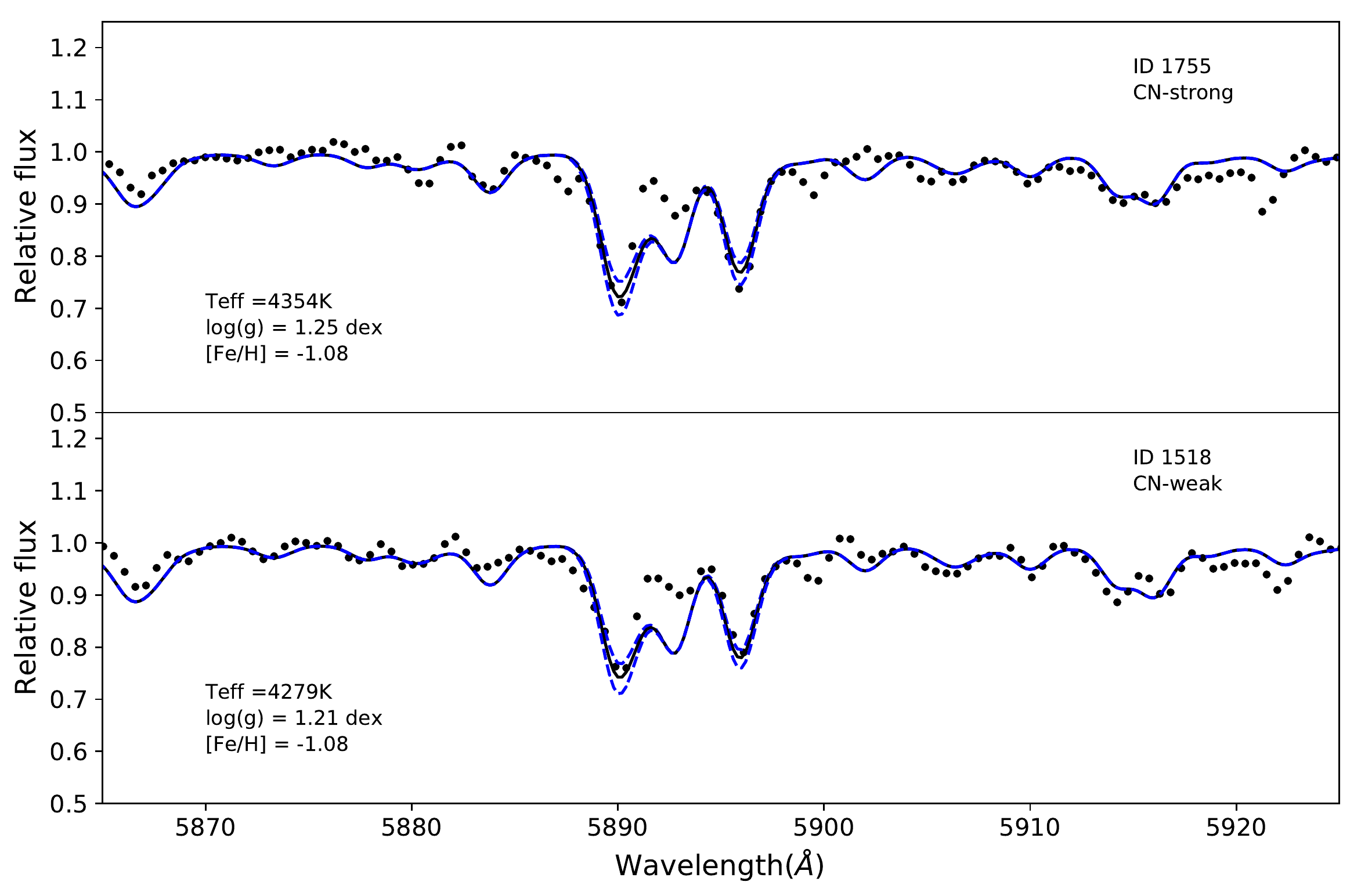}
\caption{Upper panel: Spectrum synthesis of the NaD features for star ID~1755 (CN-strong).   The solid line represents the best abundance fit  which has [Na/Fe] = --0.05 dex.  The dashed blue lines show [Na/Fe] values $\pm$ 0.15 dex about the central value. Lower panel: Spectrum synthesis of NaD features for star ID~1518 (CN-weak). The solid line represents the best abundance fit  which has [Na/Fe] = --0.32 dex. The dashed blue lines show [Na/Fe] values $\pm$ 0.15 dex about the central value. The continuum normalized  spectrum is represented by solid dots. Note that for an unknown reason the Ni{\sc i} line at $\lambda$5892.8 \AA \ is poorly reproduced in the synthetic spectra but the fits to the NaD lines are unaffected.}
\label{synth_example_na}
\end{center}
\end{figure*}

To quantify and give an interpretation of Fig.\ \ref{wnad_smc}, we have performed spectrum synthesis calculations to determine Na abundances using the same atmospheric parameters as for the spectrum synthesis of the blue spectra. An example of this procedure is shown in Fig.\ \ref{synth_example_na} for the CN-strong star (ID 1755) and the CN-weak star (ID 1518).  With the exception of the Ni{\sc i} line at $\lambda$5892.8 \AA, which for some unknown reason is substantially stronger in the synthetic spectra than in the observed stars, the fit to the NaD lines and other features in the adopted wavelength range is good.  The total uncertainty in the derived [Na/Fe] values, $\sigma_{total}$, is, as for the blue spectra, a combination of the errors from the uncertainties in the stellar parameters, $\sigma_{SP}$, and the uncertainty in the fitted value,  $\sigma_{fit}$.  The values of $\sigma_{total}$ are given in Table \ref{tabla_total_abund_SMC}.

\begin{figure}
\begin{center}
\includegraphics[width=8.5cm]{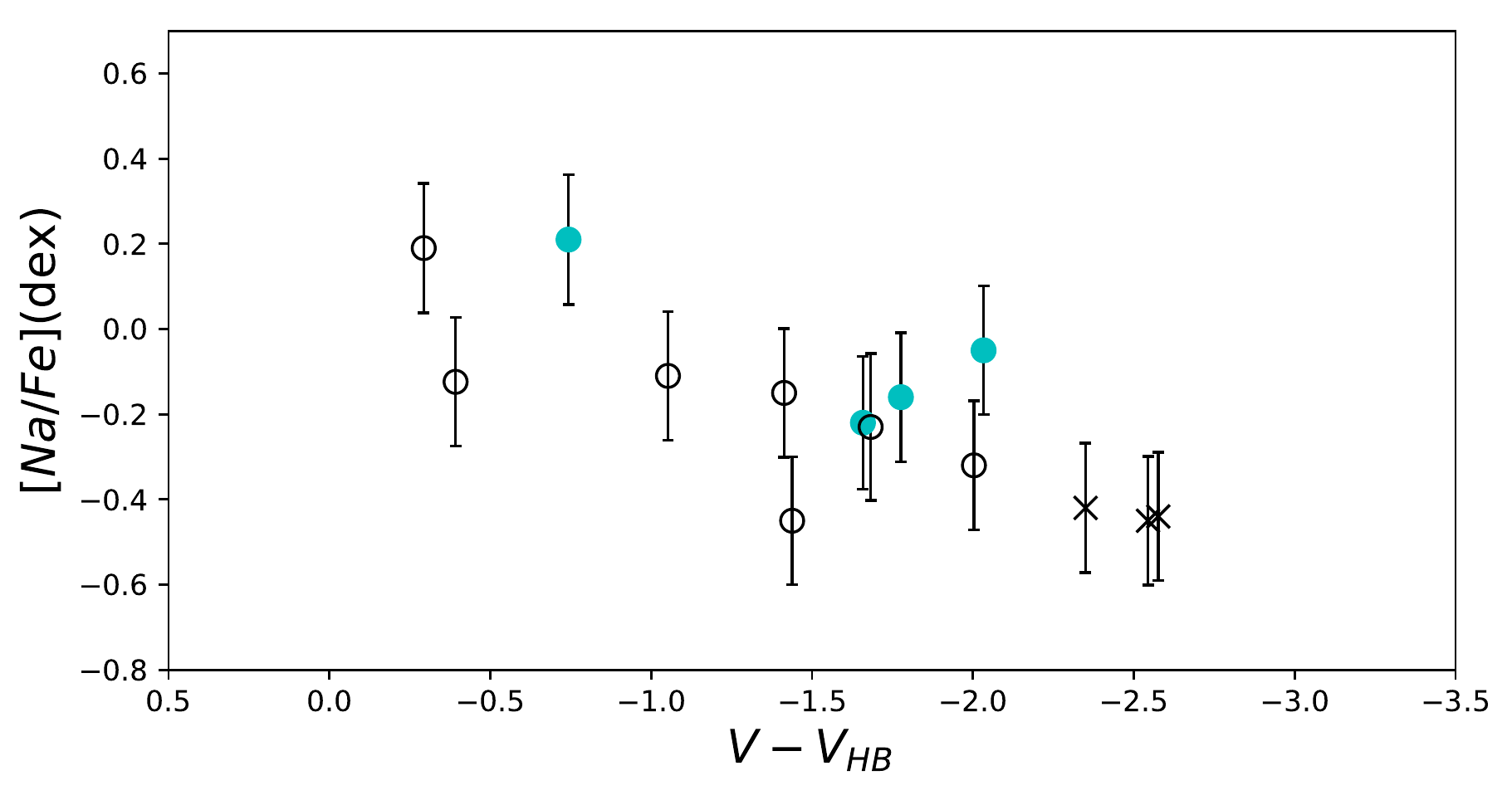}
 \caption{The dependence of [Na/Fe] on $V-V_{HB}$ magnitude for RGB members of Kron~3. Filled and empty circles  represent CN-strong and  CN-weak stars as defined by Fig.\ \ref{fig_panel_SMC_3839}. Individual uncertainties are shown for each star. The x-symbols are the three cooler and more luminous CN-weak stars, which appear to have systematically lower [Na/Fe] values. }
\label{na_smc}
\end{center}
\end{figure}

The full set of derived [Na/Fe] abundances are plotted against $V-V_{HB}$ in Fig.\ \ref{na_smc}.   It is evident in this Figure that there is some dependence of the derived [Na/Fe] values on luminosity, in the sense that the three most luminous stars have lower [Na/Fe] values than the rest of the sample.  Since we do not expect there to be any genuine change in [Na/Fe] with luminosity for the CN-weak stars (Na is not made in RGB evolutionary mixing processes), the offset likely reflects a systematic issue with the derived [Na/Fe] values for the most luminous stars.  We suggest that the offset is a consequence of the process used to perform the continuum normalization: cooler and more luminous stars have increased line-blanketing and this may have resulted in the pseudo-continuum being set too low, causing a reduction in the derived [Na/Fe] abundances. There are, however, no CN-strong stars with [Na/Fe] values at these luminosities, so we can afford to neglect these three stars in the subsequent discussion, relying instead on the $V-V_{HB}$ range that contains [Na/Fe] determinations for both CN-weak and CN-strong stars.  For completeness, we note that we have paid close attention to the observed spectrum and NaD line spectral fits for the star ID 2054, which has the lowest [Na/Fe] value in Fig.\ \ref{na_smc}.  We find no reason to question the derived [Na/Fe] value.

For the 7 CN-weak stars that overlap in $V-V_{HB}$ with the CN-strong stars, the mean value of [Na/Fe] is --0.17, with a standard deviation of the mean of 0.08 dex.  Similarly, for the 4 CN-strong stars, the mean [Na/Fe] is --0.055 with a standard error of the mean of 0.10 dex.  The CN-strong stars therefore have a higher mean [Na/Fe] abundance by $\Delta<$[Na/Fe]$>$ = 0.12 $\pm$ 0.12, which although only 1.0$\times$ the combined uncertainty in the means, does support higher Na abundances in the CN-strong stars.  Based on Fig. \ref{na_smc} (but excluding the most luminous stars), the observed range in [Na/Fe] is approximately 0.6 dex, from --0.4 to +0.2 dex.  This is consistent with the results of \citet{salgado2019investigation} who used a similar technique (but with somewhat higher resolution spectra) to determine a [Na/Fe] range of $\sim$0.5 dex in the GGC NGC~288.  This GGC has a similar overall metallicity to Kron 3.

We have investigated whether the observed difference in mean [Na/Fe] between the CN-strong and the CN-weak stars might have resulted solely from a statistical fluctuation, as follows.  We adopt the null hyphothesis that there is no intrinsic [Na/Fe] difference, i.e., that all the stars have the same [Na/Fe].  Then assuming the 1$\sigma$ error in the measured [Na/Fe] values is 0.15 dex (see Table \ref{tabla_total_abund_SMC}), we can conduct trials in which the assumed constant [Na/Fe] value for each star is perturbed by gaussian distributed errors.  We can then ask the question: for a large number of trials, how frequent is the occurrence of a subset of 4 stars having a mean [Na/Fe] equal to, or exceeding, the mean [Na/Fe] of the remaining 7 stars by the observed value of 0.12 dex?  We find from 10,000 trials that this occurs in only 11\% of the cases, which supports the reality of the observed mean [Na/Fe] difference. Further, if we increase assumed abundance error to 0.2 dex, which is the standard deviation in the [Na/Fe] values for CN-weak stars, then the observed mean [Na/Fe] abundance difference still only occurs in $\sim$18\% of the trials, again arguing for the occurrence of a real difference in mean [Na/Fe] values between the CN-strong and CN-weak smaples.

\begin{figure}
\begin{center}
\includegraphics[width=8.5cm]{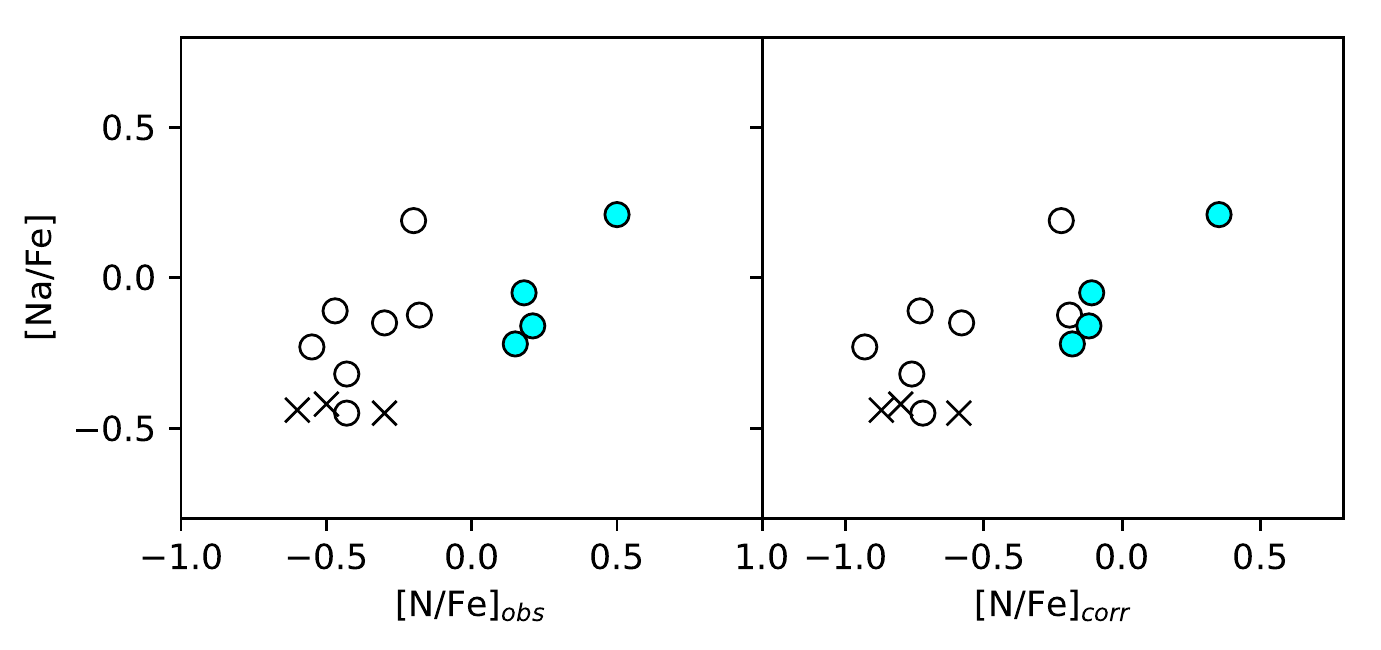}
\caption{The dependence of [Na/Fe] on [N/Fe] for RGB members of Kron~3.
The left panel shows the [N/Fe] values before the evolutionary mixing correction, while the right panel shows the corrected values. CN-strong and CN-weak stars are represented by filled and open circles,
respectively. The x-symbols represent the three cooler and more luminous CN-weak stars whose [Na/Fe] values may have been underestimated.}
\label{Na_N_panels}
\end{center}
\end{figure}

Figure \ref{Na_N_panels} shows the values of [Na/Fe] as a function of [N/Fe] for the CN-weak and CN-strong stars that overlap in $V-V_{HB}$.  The left panel employs the observed [N/Fe] values, while the right panel uses the [N/Fe] values after the application of the evolutionary mixing corrections.  In both panels the CN-strong stars have both higher [N/Fe] and higher [Na/Fe] compared to the CN-weak population, analogous to what is seen in GGCs: {\it the result argues strongly for the presence of multiple populations in Kron 3}, an intermediate-age SMC star cluster.  We note further that if the [N/Fe] axis in the panels of Fig \ref{Na_N_panels} is reversed, then the distribution of points bears a strong resemblance to the classic [Na/Fe] vs [O/Fe] anti-correlation plot seen for GGCs (e.g., \citet{carretta2009anticorrelationGiraffe}, Fig. 6).

\section{Discussion}
\label{disc_3}

\begin{figure*}
\begin{center}
\includegraphics[width=15cm]{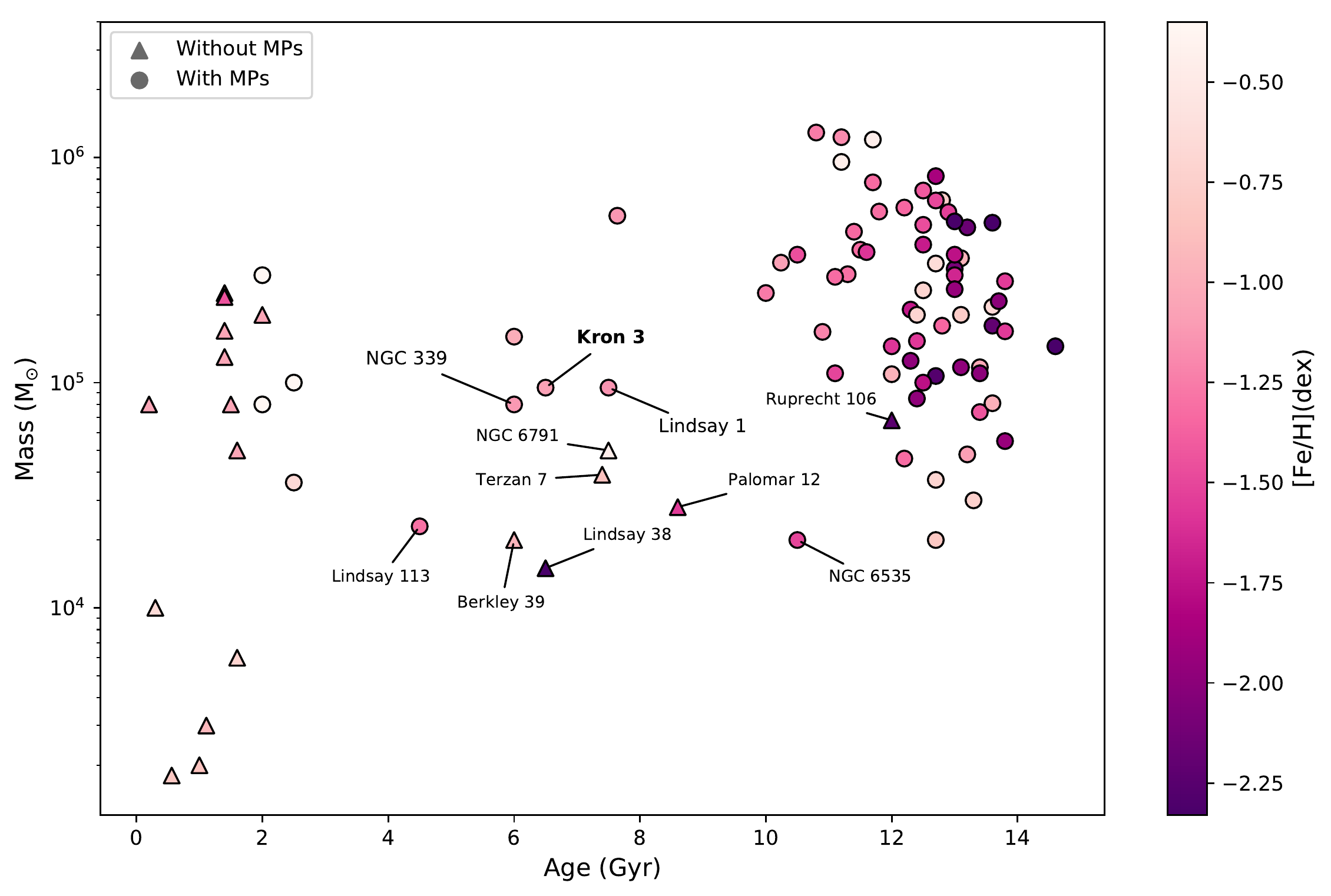}
\caption[Cluster mass versus cluster age diagram.]{Cluster mass versus cluster age diagram. Circles denote clusters containing multiple populations and triangles represent clusters where multiple populations have not been detected. Kron~3 is highlighted in bold and is plotted at the mass given by \citet{hollyhead2018kron}. The data are taken from: \citet{hollyhead2016evidence},  \citet{hollyhead2018kron}, \citet{hollyhead2019spectroscopic},  \citet{krause2016gas}, \citet{martocchia2017search},\citet{martocchia2018search}, \citet{martocchia2019search}, \citet{niederhofer2016search}.}
\label{MP_NMP}
\end{center}
\end{figure*}

The discovery of correlated variations in sodium and nitrogen abundances in a sample of individual red giants in the 6.5 Gyr old SMC cluster Kron 3 is the first direct star-by-star confirmation of the similarity between these abundance anomalies and those present in the multiple populations found in the ancient GGCs.  The importance of detecting correlated Na/N abundances to confirm the presence of multiple populations in intermediate-age star clusters has also been highlighted in the recent work of \citet{saracino2020leveraging} on the $\sim$2 Gyr old LMC cluster NGC 1978, and of \citet{martocchia2020leveraging} on the SMC clusters NGC 416 (age $\sim$6.5 Gyr) and Lindsay 1 (age $\sim$7.5 Gyr).  These studies used multi-colour HST photometry to separate red giant branch stars into two populations with different N content.  VLT/MUSE spectra of stars from each population were then combined to produce high S/N representative spectra and, in an analogous approach to that followed here, spectrum synthesis calculations for the Na D lines were then employed to estimate the [Na/Fe] abundance difference between the two populations.  In all three clusters the population that is enhanced in N is also enhanced in Na. Specifically, \citet{saracino2020leveraging}  find a [Na/Fe] abundance difference of 0.07 $\pm$ 0.01 dex for NGC 1978, while \citet{martocchia2020leveraging} find $\Delta<$[Na/Fe]$>$ = 0.18 $\pm$ 0.04 for NGC 416 and 0.24 $\pm$ 0.05 for Lindsay 1.  Since the approach of combining individual MUSE spectra effectively provides an estimate of the mean abundance difference between the two populations, the point of comparison with our work is then the mean difference in [Na/Fe].  We find for Kron 3 $\Delta <$[Na/Fe]$>$ = 0.12 $\pm$ 0.12 with the Na-rich stars also being enhanced in N, a value in reasonable accord with the estimates of \citet{martocchia2020leveraging} for the SMC clusters of similar age to Kron 3.

As stressed by  \citet{saracino2020leveraging} and \citet{martocchia2020leveraging} the existence of correlated Na and N abundance variations, as also found here for Kron 3, is compelling evidence that the same process responsible for generating the multiple populations in the ancient GGCs also occurs in intermediate-age star clusters.  It appears very likely then that whatever the process is, it cannot only be an early cosmological effect that occurs solely in ancient GGCs.

We can put our Kron 3 results in a broader context by following \citet{hollyhead2018kron} and \citet{2020IAUS..351..329M} --- see also \citet{carretta2010properties} --- in constructing a cluster mass versus age diagram for a variety of star clusters, from the old GGCs to the $\sim$1 Gyr old star clusters in the Magellanic Clouds, highlighting the clusters that do (and don't) show evidence for the presence of multiple populations.  The diagram is shown in Fig.\ \ref{MP_NMP}, and we draw attention to a few particular clusters, including Kron 3\footnote{\small{For consistency we have used the Kron 3 mass of 0.95 $\times 10^5$M$_\odot$ given by Hollyhead et al. (2018).}}. As has been noted before (e.g., \citet{2020IAUS..351..329M}, \citet{martocchia2021nitrogen}), Fig.\ \ref{MP_NMP} reveals that every cluster studied older than $\sim$2 Gyr and with a (present-day) mass exceeding $\sim10^5$M$_\odot$ shows evidence for multiple populations, while the lowest mass clusters with MPs (Lindsay 113, \citet{martocchia2019search} and NGC 6535, \citet{bragaglia2017ngc}) have present-day   masses of $\sim$2$\times 10^4$M$_\odot$. 

Five clusters with ages between 6 and $\sim$9 Gyr, however, lack evidence for MPs despite having present-day masses of, or exceeding, $\sim$2 $\times 10^4$M$_\odot$.  These are NGC 6791 \citep{villanova2018ngc}, Terzan 7 \citep{lagioia2019role}, Berkley 39 \citep{bragaglia2012searching}, Lindsay 38 \citep{martocchia2019search} and Pal 12 \citep{cohen2004palomar, kayser2008comparing}, but see also \citet{pancino2010}.  For mass to be the major discriminant governing the presence or absence of multiple populations, it would seem necessary for NGC 6535 and Lindsay 113 (L113) to have lost considerable amounts of mass.  While this is plausible for NGC 6535 given its location in the central regions of the Galaxy, considerable mass loss would appear unlikely for L113 given its location in the outer parts of the SMC. Assuming the L113 results are valid (they are based on analysis of U-B photometry rather than a full Chromosome-Map analysis), it would appear that at least one other controlling factor beyond mass maybe required.  Spectroscopic confirmation of the presence of the abundance anomalies in L113 is highly desireable. 

Further, the apparent cutoff of the presence of multiple populations at an age of $\sim$2 Gyr (e.g., see \citet{martocchia2021nitrogen}, which is set by the presence of N-variations in NGC 1978 and NGC 1651, both with age $\sim$2 Gyr, but which are absent among the RGB stars in the $\sim$1.7 Gyr old cluster NGC 1783, although N-variations may be present in the unevolved main sequence star population \citep{cadelano2021expanding}, remains an intriguing and unsolved puzzle.  This is particularly intriguing as the masses of at least some of the 1-2 Gyr old LMC clusters that apparently lack multiple populations will remain above $10^5$M$_\odot$ when they are 2-4 Gyr older. \citet{milone2020multiple} also find that the clusters NGC 419, 1783, 1806 and 1846, all of which have ages of $\sim$1.6 Gyr and present-day masses above 10$^{5}$M$_\odot$, also lack any evidence for MPs.  In contrast, the presence of multiple pops is confirmed in the slightly older cluster NGC~1978 \citep{milone2020multiple}, although the 1P fraction, at 0.85 $\pm$ 0.04\footnote{In contrast, the recent results of \citet{Li2021msstars} for main sequence stars in NGC~1978 suggest a N-rich population of at least 40\%, corresponding to a 1P fraction of $\leq$0.6.}, is substantially larger than those of GCs of similar present-day mass, further increasing the questions about these younger clusters.

\begin{figure}
\begin{center}
\includegraphics[width=8.5cm]{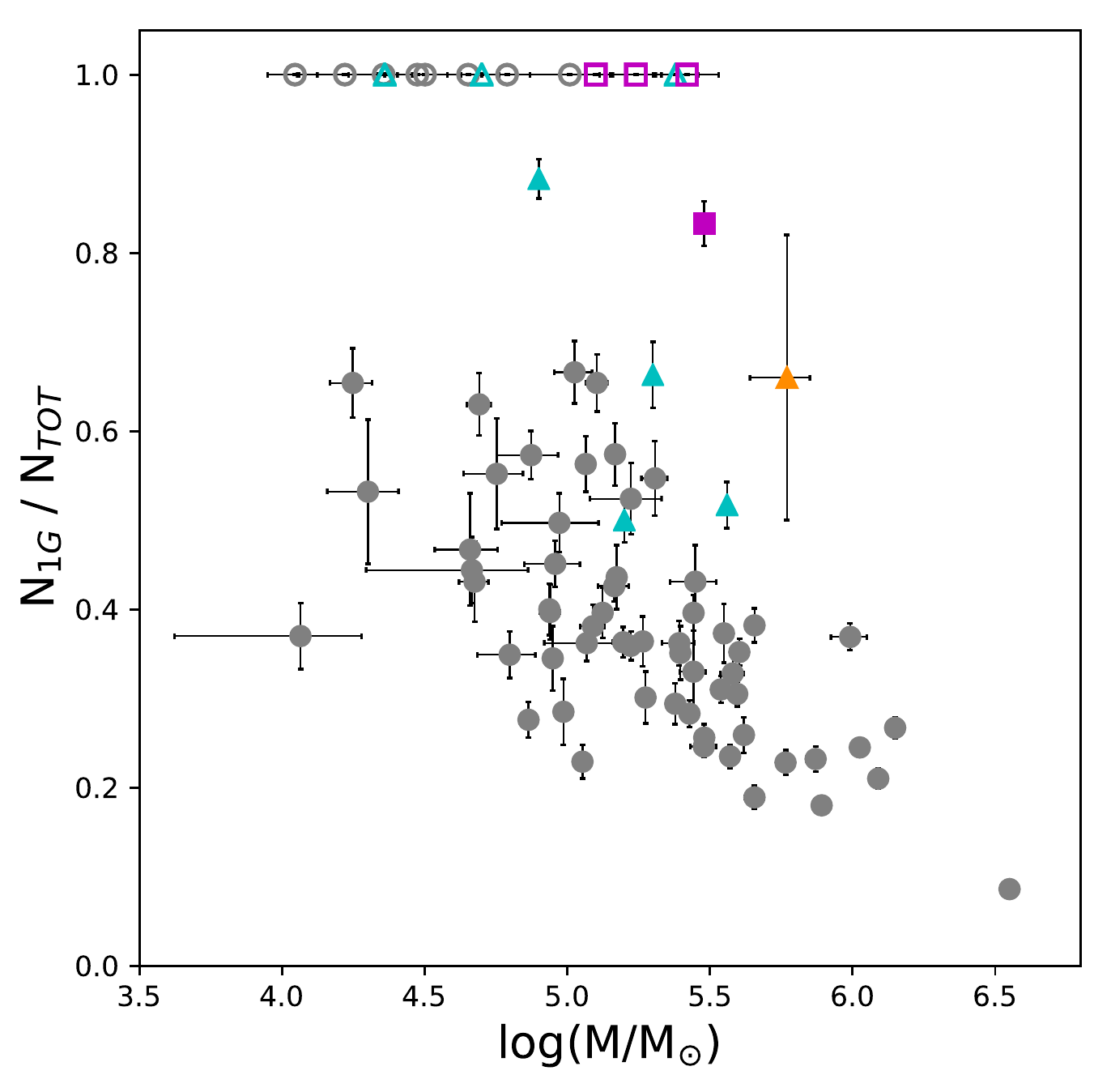}
\caption{Fraction of 1G stars against the present-day cluster mass. Galactic GCs are represented as grey dots, LMC clusters as magenta squares, and SMC clusters as cyan triangles. Kron 3 is highlighted as orange triangle. Open symbols with N$_{1G}$/N$_{tot}$ = 1.0 indicate clusters with no evidence of MPs.  Except from Kron 3, data are from \citet{baumgardt2018catalogue},  \citet{milone2020multiple} and \citet{dondoglio2021multiple}.}
\label{fig7_AM}
\end{center}
\end{figure}

We can use our results and those of others to further investigate the role of cluster mass in determining the relative fraction of primordial and second generation stars in clusters with multiple populations.  For the GGCs, based on the HST  Survey of 59 Galactic Globular Clusters, \citet{milone2020multiple} argued that the fraction of second generation stars in a GGC correlates with the cluster mass: more massive clusters have a larger second generation fraction (or equivalently, a smaller first generation fraction).  In Fig. \ref{fig7_AM}, we show the data from \citet{baumgardt2018catalogue}, \citet{milone2020multiple} and \citet{dondoglio2021multiple} including the results for LMC and SMC clusters.  We note that the 1P fractions of \citet{milone2020multiple} and \citet{dondoglio2021multiple} for the LMC and SMC clusters are generally consistent with those from other studies.  For example, 
\citet{hollyhead2016evidence} and \citet{niederhofer2016search} list 0.68 for Lindsay~1 while \citet{milone2020multiple} give 0.66 $\pm$ 0.04.
For NGC 339, NGC 416 and NGC 121 values are 0.75 \citep{niederhofer2016search}, 0.55 \citep{niederhofer2016search} and 0.68 
\citet{niederhofer2016search2}, respectively, while the \citet{milone2020multiple} values are 0.88 $\pm$ 0.02, 0.48 $\pm$ 0.03 and 0.52 $\pm$ 0.03, respectively.
Fig. \ref{fig7_AM} shows that the 1P fraction for GGCs with (present-day) masses $\sim10^5$ M$_\odot$ ranges between 
$\sim$0.2 and 0.65, and there are only two clusters in the sample (NGC~339 and NGC~1978) that have a P1 fraction above 0.7, aside from the 4 clusters in the LMC/SMC with ages less than 2 Gyr that lack MPs (1P fraction = 1.0). In the \citet{milone2020multiple} analysis NGC 339 has a present-day mass of $8\times10^{4}$M$_\odot$ while NGC 1978 has a mass of $3\times10^{5}$M$_\odot$.  

 We can now add Kron 3 into the diagram.  We have a P1 fraction of 0.67 $\pm$ 0.16, in good agreement with the value 0.69 from \citet{hollyhead2018kron}).  We adopt a Kron 3 (present-day) mass of $5.8\times10^{5}$M$_\odot$ from \citet{glatt2011present} as this seems the most consistent choice with the cluster masses given in \citet{milone2020multiple}
 \footnote{The cluster masses used in \citet{milone2020multiple} are based on Single Stellar Population (SSP) models.  The recent work of \citet{song2021dynamical} has found that their dynamically determined cluster masses are typically 40\% lower than the SSP-based masses.}.  The location of Kron 3 in Fig. \ref{fig7_AM} is substantially above the 1G fractions for GGCs of similar present-day masses.  Given Kron 3's location in the SMC, dynamical effects are likely minimal and, given the $\sim$6.5 Gyr cluster age, the mass of Kron 3 at an age of $\sim$13.5 Gyr is likely to be only slightly less than its current value -- we estimate a change of less than 10\% as a result of stellar evolution.  The same argument likely also applies to NGC~339: both clusters at the GGC age would still lie above the GGC relation.  Of course what matters is the dynamical environment, if Kron 3 and NGC 339 lived their lives in the central regions of the Milky Way, their present-day masses would likely be lower.  Taken together these results therefore indicate that while mass is likely the major factor governing the relative numbers of primordial and second generation stars in star clusters that have multiple populations, environment and age may also play a role.

\section{Summary}
\label{concl_3}

Spectra obtained with the VLT/FORS2 and Gemini-S/GMOS-S instruments of red giant stars in the $\sim$6.5 Gyr old SMC star cluster Kron 3 have yielded 18 likely cluster members. The membership selection process, detailed in Section \ref{membership_3}, was extensive and we are confident of the assigned membership status.  The highlights of this work can be summarized as:

\begin{itemize}

\item In Kron~3, 6 of the 18 members show strong CN-band strengths. Plots of the CN-band strength indices S(3839) and S(4142) against $V-V_{HB}$ (see Figs. \ref{fig_panel_SMC_3839} and  \ref{fig_panel_SMC_4142_wg}) indicate a significant observed spread and allow the definition of CN-strong and CN-weak stars.  The population fraction [N(CN-strong)/(N(CN-weak)+N(CN-strong))] is 0.33 $\pm$ 0.16.

\item The plot of the CH-band strength index W(G) against $V-V_{HB}$ (Fig. \ref{fig_panel_SMC_4142_wg}) does not show a significant dispersion.  However, on average, the CN-strong stars have smaller W(G) values compared the CN-weak stars at similar $V-V_{HB}$ magnitude.  As shown in Fig. \ref{correlation_SMC}, a CN/CH anti-correlation is present, a likely indication of the presence of multiple populations in Kron 3 similar to those seen in GGCs.

\item Spectrum synthesis calculations have been used to derive carbon and nitrogen abundances from the observed spectra.  As shown in Fig. \ref{evolutionary_mixing}, there is a marked decrease in the [C/Fe] values with increasing luminosity, which we ascribe to evolutionary mixing on the RGB.  Overall, there is an anti-correlation between the C and N abundances.  The observed range in [N/Fe] is $\sim$1.4 dex and the observed range in [C/Fe], after applying corrections for evolutionary mixing, is 
$\sim$0.5 dex.  Both values agree well with those found for Kron 3 in the independent study of \citet{hollyhead2018kron}.

\item Investigation of the Na~D lines shows that the CN-strong stars generally have stronger W(NaD) values compared to CN-weak stars at similar luminosities (Fig. \ref{wnad_smc}).  Applying spectrum synthesis calculations reveals an observed  [Na/Fe] abundance range of $\sim$0.6 dex with, on average, the CN-strong stars having higher [Na/Fe] compared to the CN-weak stars by $\sim$0.12 dex.  Significantly, there is a distinct Na/N correlation: the CN-strong stars have both higher [Na/Fe] and higher [N/Fe] compared to the CN-weak population.  This relation argues strongly for the presence of multiple populations in Kron 3 analogous to those seen in the ancient GGCs.

\item In Kron 3, and in other intermediate age star clusters, the relative population fraction of the second generation stars (e.g., [N(CN-strong)/(N(CN-weak)+N(CN-strong))]) is lower than it is for GGCs of similar present-day mass and luminosity.  This hints that cluster mass is not the only determinant of the population fraction, age and environment may also play a role.

\end{itemize}

Overall, we have demonstrated that the multiple population phenomenon, prevalent in the ancient GGCs, is also present in the 6.5 Gyr old SMC cluster Kron 3.  This confirms that the mechanism responsible for the multiple populations is not restricted just to the oldest clusters: younger clusters with sufficient mass (and seemingly with an age exceeding $\sim$2 Gyr) also possess multiple populations.

\newpage
\begin{table*}
\caption[Observational data for member stars in Kron~3 observed with FORS2 and GMOS-S.]{Observational data for member stars in Kron~3 observed with FORS2 and GMOS-S. Some S(3839) values are missing due as a result of reduced wavelength coverage. Missing W(NaD)  values result from low S/N for the fainter stars or because they were not observed on the GMOS-S run. Empty circles denote the stars classified as CN-weak while filled circles represent the CN-strong stars.}

\resizebox{\linewidth}{!}
{\begin{threeparttable}
\renewcommand{\TPTminimum}{\linewidth}
\centering 
\begin{adjustbox}{max width=15cm}
\begin{tabular}{l c c c c c c c c c c c c c c r}
\hline 
\hline

\normalsize{Cluster} & \normalsize{ID} & \normalsize{RA}      & \normalsize{Dec} & \normalsize{V}   & \normalsize{B-V} &  
\normalsize{RV} & \normalsize{Distance$^1$} &
\normalsize{S(3839)} & \normalsize{S(4142) } &\normalsize{W(G) } & \normalsize{W(NaD)} & \\ [0.5ex]
\normalsize{ }  & \normalsize{ }  & \normalsize{J2000} & \normalsize{J2000} & \normalsize{mag} & \normalsize{mag} &  \normalsize{km s$^{-1}$} & \normalsize{arc sec} & \normalsize{mag}     & \normalsize{mag}      &\normalsize{\AA}   & \normalsize{\AA}   & 	& \\

\hline 

Kron~3 &	2237   &   0:24:50.02   &   -72:48:35.4   &   16.837   &   1.395    &    127.1	&  95.4 	& 0.212	&   -0.099	&   11.35	&   -		& $\circ$  &\\  
Kron~3 &	1658   &   0:24:46.86   &   -72:46:46.4   &   16.841   &   1.460    &    141.9	&  55.0 	& 0.391	&   -0.071	&   11.18	&   -		& $\bullet$ & \\
Kron~3 &	1372   &   0:24:34.98   &   -72:47:28.8   &   16.924   &   1.453    &    129.6	&  152.1 	& 0.134	&   -0.181	&   11.23	&   1.096	& $\circ$  & \\
Kron~3 &	1712   &   0:24:35.64   &   -72:46:22.0   &   16.956   &   1.395    &    129.8	&  159.7 	& 0.160	&   -0.182	&   11.49	&   0.732	& $\circ$  &\\
Kron~3 &	1096   &   0:24:35.60   &   -72:47:56.0   &   17.150   &   1.331    &    129.4	&  144.3 	& 0.159	&   -0.197	&   11.33	&   0.886	& $\circ$  &\\
Kron~3 &	1755   &   0:24:50.83   &   -72:45:54.9   &   17.467   &   1.159    &    131.0	&  131.5 	& 0.543	&   -0.156	&   10.48	&   1.097	& $\bullet$ & \\
Kron~3 &	1518   &   0:24:37.84   &   -72:47:12.3   &   17.497   &   1.214    &    146.0	&  111.4 	& 0.226	&   -0.227	&   11.39	&   0.772	& $\circ$  &\\
Kron~3 &	1659   &   0:24:51.90   &   -72:46:45.9   &   17.724   &   1.140    &    127.3	&  112.8 	& 0.412	&   -0.204	&   11.27	&   0.893	& $\bullet$ & \\
Kron~3 &	1262   &   0:24:37.12   &   -72:47:37.3   &   17.818   &   1.093    &    144.8	&  119.9 	& 0.044	&   -0.284	&   11.73	&   0.902	& $\circ$  &\\
Kron~3 &	1564   &   0:24:40.31   &   -72:47:06.5   &   17.842   &   1.112    &    131.1	&  77.4 	& 0.481	&   -0.197	&   10.52	&   1.123	& $\bullet$ & \\
Kron~3 &	2054   &   0:24:33.90   &   -72:50:20.5   &   18.062   &   0.978    &    123.5	&  236.1 	& 0.026	&   -0.282	&   10.39	&   0.642	& $\circ$  &\\
Kron~3 &	1736   &   0:24:51.91   &   -72:46:05.0   &   18.087   &   0.973    &    145.4	&  135.7 	& 0.062	&   -0.296	&   10.17	&   0.677	& $\circ$  & \\
Kron~3 &	1688   &   0:24:43.60   &   -72:46:34.2   &   18.448   &   0.976    &    147.0	&  64.7 	& 0.058	&   -0.306	&   10.96	&   0.884	& $\circ$  & \\
Kron~3 &	2301   &   0:24:34.96   &   -72:48:16.9   &   18.757   &   0.874    &    126.1	&  158.1 	& 0.368	&   -0.249	&    9.34	&   1.124	& $\bullet$ & \\
Kron~3 &	2161   &   0:24:39.76   &   -72:49:11.7   &   19.108   &   0.823    &    129.6	&  125.9 	& -0.146&   -0.336	&   10.64	&   0.657	& $\circ$  &\\
Kron~3 &	2086   &   0:24:36.42   &   -72:49:57.7   &   19.207   &   0.813    &    134.6	&  193.5 	& -0.160&   -0.337	&    9.76	&   0.785	& $\circ$  &\\
Kron~3 &	2044   &   0:24:49.25   &   -72:50:38.2   &   19.264   &   0.797    &    129.2	&  193.6 	& 0.278	&   -0.278	&    9.12	&   -		& $\bullet$ &\\
Kron~3 &	1876   &   0:24:47.09   &   -72:44:13.2   &   19.390   &   0.804    &    130.2	&  203.8 	& -0.039&   -0.310	&   10.54	&   -		& $\circ$  &\\
\\

\hline 
\end{tabular}
\end{adjustbox}
 

\end{threeparttable}}
\begin{tablenotes}
\item [1] (1) Distance corresponds to the distance from the cluster centre.

\end{tablenotes}
\label{tabla_total_index_SMC}
\end{table*}

\begin{table*}
\caption[Observational data for non-member stars in Kron~3.]{Observational data for non-member stars the vicinity of Kron~3.}

\resizebox{\linewidth}{!}
{\begin{threeparttable}
\renewcommand{\TPTminimum}{\linewidth}
\centering 
\begin{adjustbox}{max width=15cm}
\begin{tabular}{l c c c c c c c c c r}
\hline 
\hline

\normalsize{Cluster} & \normalsize{ID} & \normalsize{RA}      & \normalsize{Dec} & \normalsize{V}   & \normalsize{B-V} &  
\normalsize{RV} & \normalsize{Distance$^1$} & \\ [0.5ex]

\normalsize{ }  & \normalsize{ }  & \normalsize{J2000} & \normalsize{J2000} & \normalsize{mag} & \normalsize{mag} &  
\normalsize{km s$^{-1}$} & \normalsize{arc sec}  & \\

\hline 
		
Kron~3 &	2177 &	0:24:36.00 	&	-72:48:57.6 &	16.832 	& 1.348	& 124.1 	&  160.3	&\\ 
Kron~3 &	2143  &    0:25:05.19         &       -72:49:19.7 &    17.242      & 1.163     & 120.8	&  318.9 	&\\
Kron~3 &	2286 &	0:24:47.04 	&	-72:48:21.6 &	17.355 	& 1.209	& 107.1 	&  53.4	&\\ 
Kron~3 &	2240 &	0:24:47.52 	&	-72:48:36.0 &	17.942 	& 1.075	& 157.9 	&  70.3	&\\ 
Kron~3 &	1123 &	0:24:43.20 	&	-72:47:52.8 &	18.187 	& 1.010	& 152.5	&  33.9	&\\
Kron~3 &	2020 &	0:24:17.52 	&	-72:50:56.4 &	18.514 	& 0.941	& 173.0 	&  460.2	&\\ 
Kron~3 &	1792 &	0:25:12.72 	&	-72:45:32.4 &	18.650 	& 0.887    & 109.5	&  432.5 	&\\  
Kron~3 &	2069  &     0:24:22.08         &      -72:50:09.4 &    18.982      & 0.830    & 131.2	&  378.5 	&\\
Kron~3 &	1830   &    0:24:24.87         &      -72:44:59.1 &    19.037      &  0.941   & 119.7	&  341.3 	&\\
Kron~3 &	1859 &	0:24:11.04 	&	-72:44:31.2 &	19.199 	& 0.718	& 165.5 	&  544.6	&\\ 
Kron~3 &	2041 &	0:24:13.92 	&	-72:50:42.0 &	19.242 	& 0.776	& 148.4 	&  505.5	&\\ 
Kron~3 &	1798   &    0:24:22.48         &      -72:45:30.1 &    19.243      & 0.882    &  144.2	&  361.6 	&\\
Kron~3 &	2110 &	0:24:21.60 	&	-72:49:40.8 &	19.283 	& 0.833	& 157.0 	&  376.1	&\\

\hline 
\end{tabular}
\end{adjustbox}

\end{threeparttable}}
\begin{tablenotes}
\item [1] (1) Distance corresponds to the distance from the cluster centre.

\end{tablenotes}
\label{tabla_total_index_SMC_nonmember}
\end{table*}

\begin{table*}
\caption{Stellar parameters and derived C, N and Na abundances, and their uncertainties, for the Kron~3 members.}
\resizebox{\linewidth}{!}
{\begin{threeparttable}
\renewcommand{\TPTminimum}{\linewidth}
\centering 
\begin{adjustbox}{max width=15cm}
\begin{tabular}{l c c c c c c c c c c c c c c r}
\hline 
\hline

\normalsize{Cluster} & \normalsize{ID} & \normalsize{V}      & \normalsize{T$_{eff}$} & \normalsize{log($g$)}   & \normalsize{[C/Fe]} &  \normalsize{e[C/Fe]} & \normalsize{[C/Fe]$_{cor}$} &\normalsize{[N/Fe]} & \normalsize{e[N/Fe]} & \normalsize{[N/Fe]$_{cor}$}  & \normalsize{[Na/Fe]} & \normalsize{e[Na/Fe]} &  \\ [0.5ex]

\normalsize{ }  & \normalsize{ }  & \normalsize{mag} & \normalsize{K} & \normalsize{dex} & \normalsize{dex} & \normalsize{dex}   & \normalsize{dex}   & \normalsize{dex}   & \normalsize{dex} & \normalsize{dex} & \normalsize{dex} & \normalsize{dex}   & \\

\hline 

Kron~3	&	2237	&	16.84	&	4050	&	0.76	&	-1.00	&	0.18	&	-0.39	&	-0.60	&	0.34	&	-0.95	&	-	  	&	-		& \\	
Kron~3	&	1658	&	16.84	&	3973	&	0.69	&	-1.10	&	0.18	&	-0.49	&	0.00	&	0.34	&	-0.29	&	-		&	-		& \\	
Kron~3	&	1372	&	16.92	&	3981	&	0.73	&	-1.15	&	0.18	&	-0.54	&	-0.60	&	0.34	&	-0.87	&	-0.44	&	0.15	& \\	
Kron~3	&	1712	&	16.96	&	4050	&	0.81	&	-1.10	&	0.18	&	-0.49	&	-0.30	&	0.34	&	-0.59	&	-0.45	&	0.15	& \\	
Kron~3	&	1096	&	17.15	&	4128	&	0.95	&	-1.08	&	0.18	&	-0.47	&	-0.50	&	0.34	&	-0.80	&	-0.42	&	0.15	& \\	
Kron~3	&	1755	&	17.47	&	4354	&	1.25	&	-1.05	&	0.17	&	-0.49	&	0.18	&	0.33	&	-0.11	&	-0.05	&	0.15	& \\	
Kron~3	&	1518	&	17.50	&	4279	&	1.21	&	-0.94	&	0.18	&	-0.40	&	-0.43	&	0.33	&	-0.76	&	-0.32	&	0.15	& \\	
Kron~3	&	1659	&	17.72	&	4381	&	1.37	&	-0.90	&	0.17	&	-0.39	&	0.21	&	0.34	&	-0.12	&	-0.16	&	0.15	& \\	
Kron~3	&	1262	&	17.82	&	4448	&	1.45	&	-0.70	&	0.18	&	-0.26	&	-0.55	&	0.33	&	-0.93	&	-0.23	&	0.17	& \\	
Kron~3	&	1564	&	17.84	&	4421	&	1.44	&	-0.89	&	0.17	&	-0.38	&	0.15	&	0.33	&	-0.18	&	-0.22	&	0.15	& \\	
Kron~3	&	2054	&	18.06	&	4622	&	1.65	&	-0.83	&	0.17	&	-0.41	&	-0.43	&	0.31	&	-0.72	&	-0.45	&	0.14	& \\	
Kron~3	&	1736	&	18.09	&	4630	&	1.67	&	-0.83	&	0.17	&	-0.43	&	-0.30	&	0.31	&	-0.58	&	-0.15	&	0.15	& \\	
Kron~3	&	1688	&	18.45	&	4625	&	1.81	&	-0.77	&	0.17	&	-0.43	&	-0.47	&	0.31	&	-0.73	&	-0.11	&	0.15	& \\	
Kron~3	&	2301	&	18.76	&	4792	&	2.02	&	-0.62	&	0.17	&	-0.46	&	0.50	&	0.32	&	0.35	&	0.21	&	0.15	& \\	
Kron~3	&	2161	&	19.11	&	4880	&	2.21	&	-0.75	&	0.16	&	-0.73	&	-0.18	&	0.30	&	-0.19	&	-0.12	&	0.15	& \\	
Kron~3	&	2086	&	19.21	&	4898	&	2.26	&	-0.63	&	0.16	&	-0.61	&	-0.20	&	0.29	&	-0.22	&	0.19	&	0.15	& \\	
Kron~3	&	2044	&	19.26	&	4927	&	2.30	&	-0.78	&	0.16	&	-0.76	&	0.48	&	0.31	&	0.47	&	-		&	-		& \\	
Kron~3	&	1876	&	19.39	&	4914	&	2.34	&	-0.63	&	0.16	&	-0.61	&	0.10	&	0.31	&	0.08	&	-		&	-		& \\

\hline 
\end{tabular}
\end{adjustbox}


\end{threeparttable}}
\label{tabla_total_abund_SMC} 
\end{table*}
\newpage

\section*{Acknowledgements}
\addcontentsline{toc}{section}{Acknowledgements}
The authors are grateful for the detailed comments of the anonymous referee on the original version of this manuscript.
Based in part on observations collected at the European Southern Observatory under ESO programme 095.D-0496.  Also based in part on observations obtained at the international Gemini Observatory, a program of NSF's NOIRLab, which is managed by the Association of Universities for Research in Astronomy (AURA) under a cooperative agreement with the National Science Foundation on behalf of the Gemini Observatory partnership: the National Science Foundation (United States), National Research Council (Canada), Agencia Nacional de Investigaci\'{o}n y Desarrollo (Chile), Ministerio de Ciencia, Tecnolog\'{i}a e Innovaci\'{o}n (Argentina), Minist\'{e}rio da Ci\^{e}ncia, Tecnologia, Inova\c{c}\~{o}es e Comunica\c{c}\~{o}es (Brazil), and Korea Astronomy and Space Science Institute (Republic of Korea).
CS acknowledges support provided by CONICYT, Chile, through its scholarships program  CONICYT-BCH/Doctorado Extranjero 2013-72140033. 
APM acknowledges support from the European Research Council (ERC) under the European Union's Horizon 2020 research innovation programme (Grant Agreement ERC-StG 2016, No 716082 'GALFOR', PI: Milone, http://progetti.dfa.unipd.it/GALFOR) and to the MIUR through the FARE project R164RM93XW SEMPLICE (PI: Milone) and the PRIN program 2017Z2HSMF (PI: Bedin).

\section*{Data availability}
The data underlying this article are available will be shared on reasonable request to the corresponding author.

\bibliographystyle{mnras}
\bibliography{phdbib}
%
%
\end{document}